\documentclass[aps,pra,superscriptaddress,showpacs,reprint]{revtex4-1}
\usepackage{graphicx}
\usepackage{amsmath}
\usepackage{subfigure}
\usepackage{bm}
\usepackage{verbatim}
\usepackage[colorlinks=true, pdfstartview=FitV,urlcolor=blue,linkcolor=orange,citecolor=cyan]{hyperref}
\usepackage{color}
\usepackage{xcolor}
\usepackage{hyperref}
\usepackage{amssymb}
\usepackage{bbm}
\usepackage{dsfont}
\usepackage{siunitx}
\usepackage{txfonts}

\newcommand{\abs}[1]{\left\lvert #1\right\rvert}

\newcommand{\average}[1]{\langle#1\rangle}

\newcommand{\TwoDMatrix}[4]{\begin{pmatrix} #1 & #2 \\ #3 & #4 \end{pmatrix}}
\usepackage{mathtools}
\newcommand{\tu}{\tilde{u}}

\DeclarePairedDelimiterX\braket[2]{\langle}{\rangle}{#1 \delimsize\vert #2}
\DeclarePairedDelimiterX\inner[2]{\langle}{\rangle}{#1,#2}

\newcommand{\AAM}{Aubry-Andre}

\newcommand{\SE}{Soukoulis-Economou}

\newcommand{\ttM}{$t_1$-$t_2$ model}
\newcommand{\tp}{t_{\perp}}
\newcommand{\avgIPR}{\langle\text{IPR}\rangle}
\newcommand{\avgNPR}{\langle\text{NPR}\rangle}

\begin{document}
\title{Mobility edge and intermediate phase in one-dimensional incommensurate lattice potentials} 

\author{Xiao Li}
\email{lixiao@umd.edu}
\affiliation{Condensed Matter Theory Center and Joint Quantum Institute, University of Maryland, College Park, MD 20742, USA}
\author{S. Das Sarma}
\affiliation{Condensed Matter Theory Center and Joint Quantum Institute, University of Maryland, College Park, MD 20742, USA}

\date{\today}

\begin{abstract}
We study theoretically the localization properties of two distinct one-dimensional quasiperiodic lattice models with a single-particle mobility edge (SPME) separating extended and localized states in the energy spectrum. 
The first one is the familiar Soukoulis-Economou trichromatic potential model with two incommensurate potentials, and the second is a system consisting of two coupled 1D {\AAM} chains each containing one incommensurate potential. 
We show that as a function of the Hamiltonian model parameters, both models have a wide single-particle intermediate phase (SPIP), defined as the regime where localized and extended single-particle states coexist in the spectrum, leading to a behavior intermediate between purely extended or purely localized when the system is dynamically quenched from a generic initial state. 
Our results thus suggest that both systems could serve as interesting experimental platforms for studying the interplay between localized and extended states, and may provide insight into the role of the coupling of small baths to localized systems. 
We also calculate the Lyapunov (or localization) exponent for several incommensurate 1D models exhibiting SPME, finding that such localization critical exponents for quasiperiodic potential induced localization are nonuniversal and depend on the microscopic details of the Hamiltonian. 
\end{abstract}

\maketitle

\section{Introduction}
In the past decade, the dynamics of isolated quantum systems has attracted considerable attention in the context of the conceptual foundation of quantum statistical mechanics. 
In particular, whether a generic isolated quantum many-body system can reach thermalization under its own dynamics has been studied extensively. 
It has been claimed~\cite{MBL01,MBL02,MBL03} that an interacting many-body system may generically fail to thermalize on its own (i.e., without any coupling to an outside bath) in the presence of strong disorder, resulting in a stable dynamical nonergodic or nonthermal phase of matter at non-zero temperatures (i.e., at finite energy densities).  
Such a perfect interacting insulator has been named `many-body localized' and the phenomenon as `many-body localization' (MBL)~\cite{MBL04,MBL05,MBL06,MBL07,MBL09,MBL08,MBL10,MBL11,MBL12,MBL13,MBL14,MBL15}. 
One hallmark of MBL is that the eigenstate thermalization hypothesis (ETH)~\cite{ETH1,ETH2} is strongly violated, leading to a manifestly non-ergodic phase in spite of the system being strongly interacting. 
On the experimental side, signatures of MBL have been observed in various quantum many-body systems that are well isolated from the environment, such as ultracold atoms~\cite{Schreiber_2015,Choi_2016,Bordia_2016} and trapped ions~\cite{Smith_2016,Wu2016}. 
Currently, MBL is one of the most active research areas in condensed matter and AMO physics.

One of the outstanding questions in MBL research is the coupling between localized and extended degrees of freedom in a many-body system. 
In general, coupling to a large external bath (much larger than the system itself, by definition) would automatically restore ergodicity since the system can thermalize through its interaction with the bath.  The question, however, is what happens if the bath is small in some well-defined sense. 
In particular, it is important to understand what happens if an MBL system is coupled to an ergodic system (i.e., the bath) of a similar size. 
Will the coupled system become many-body localized, or will the coupled system become completely ergodic? 
The key question here, which is still open, is whether an MBL system can localize the bath rather than the bath thermalizing the MBL system which is the usual situation. 
There have been many theoretical~\cite{MBL08,MBL10,MBL11,MBL12,MBL13,MBL14,MBL15,Modak2015,Baygan2015,XiaoPeng_PRL,XiaoPeng_PRB,Luitz2016,Roeck16,Torres-Herrera2017,Nag2017,Roeck2017,Luitz_2017,Ponte_2017,Hsu2018} and experimental~\cite{Lueschen2017,Bordia2017,Rubio-Abadal2018} studies in this direction recently, but many questions still remain open. 
In particular, one of the controversial issues is whether it is possible to have a many-body intermediate phase, i.e., a coexistence of MBL and ETH (ergodic) states in the same system, leading to a `nonergodic metal' phase~\cite{XiaoPeng_PRL,XiaoPeng_PRB,Hsu2018} which shares some (but, not all) properties of both ETH and MBL phases. 
Several numerical studies have shown that this scenario is possible under certain circumstances~\cite{MBL11,XiaoPeng_PRL,Modak2015,Baygan2015,XiaoPeng_PRB,Nag2017,Hsu2018}, and some of the results even suggested that a many-body mobility edge (i.e., the spectrum consists of MBL states up to a specific many-body energy density and of ETH states above that) may be present. 
On the other hand, some recent work have argued on theoretical grounds that no many-body mobility edge can exist in the thermodynamic limit due to rare-region effects in a system with truly random disorders~\cite{Roeck16,Roeck2017}. 
Note that the presence of a mobility edge allows the same system to act as a bath since at finite temperatures, the extended states will act as a bath for the localized states as they exchange energy through interactions.  
But the bath here remains the same size as the system itself even in the thermodynamic limit since the mobility edge divides a finite fraction of the states into extended states leaving the other fraction localized.  
But such an intermediate phase may also arise without any mobility edges simply as a mixture of thermal and localized states due to the tuning of some other parameter~\cite{Schecter2018}. 
The main issue is whether an intermediate phase can arise when a system is quenched due to the competition between thermal and nonergodic states.  To advance our understanding of this important question, we study two models in the current work showing an intermediate phase in the non-interacting situation.  The great merit of both systems is that they are easy to study in the laboratory using ultracold atom optical lattices, enabling future studies of the corresponding interacting system in depth.
Given the importance of the general topic of MBL, particularly, the question of the existence or not of MBL in a system containing an SPME in the corresponding noninteracting situation, we believe that studying localization properties of many systems with SPME may lead to significant conceptual advance. 

In order to gain more insights into this issue, inputs from experimental studies are vital, as probing the thermodynamic limit in numerical simulations of MBL is rather difficult, if not impossible. 
To date, all numerically studied MBL systems have rather limited system sizes ($10$-$20$ particles), because the many-body Hilbert space becomes prohibitively large for more than $20$ interacting particles. 
In contrast, ultracold atomic systems can easily host more than $100$ particles, giving rise to an order of magnitude improvement in going toward the thermodynamic limit. 
Therefore, it is highly desirable to identify and study experimentally feasible models that contain both extended and localized degrees of freedom in the noninteracting limit so that the experiment can start from a well-defined limit where such an intermediate phase definitely exists in order to investigate what happens when the interaction is turned on. 
Putting such models to test using the state-of-the-art   experimental techniques can greatly improve our understanding of quantum ergodicity in the presence of small baths since the noninteracting system here has both localized and extended states with the latter presumably acting as a bath for the former.

In this regard, theoretical models containing single-particle mobility edges (SPME) are particularly attractive. 
In such systems, the single-particle energy spectrum is divided sharply into localized and extended eigenstates by a critical energy $E_c$ (i.e., the mobility edge). 
Equivalently, we also say that a single-particle Hamiltonian is in an `intermediate phase' whenever there exists one or more mobility edges in its spectrum---we note that while SPME certainly implies an intermediate phase, the reverse is not true, i.e., an intermediate phase may very well exist without any mobility edges, e.g., a simple mixture of localized and extended states arising from two disconnected subsystems. 
One of the most familiar examples is the 3D Anderson model~\cite{Anderson1958} where, if the disorder strength is modest, extended states will appear in the middle of the spectrum while localized ones are near the spectrum edges, leading to the appearance of mobility edges. 
(For very strong disorder, the whole spectrum may be localized whereas for no disorder all eigenstates are extended band states in this case.) 
We note that in 1D (as well as 2D) there are no SPME in the Anderson model with all states being strictly localized in the presence of any random disorder. 
In 1D systems, the search for models with an SPME also has a long history~\cite{mott1990metal}, with many theoretical models being proposed and studied in the past 30$\sim$40 years~\cite{SPME01,Soukoulis_Economou,SPME02,SPME03,SPME04,SPME05,SPME06,SPME07,SPME08,SPME09}, all of these models involving some types of `correlated disorder' in the form of deterministic quasiperiodic potentials. 
However, due to {experimental challenges}, the first observation of SPME in 1D systems~\cite{Xiao17,LuschenSPME17} has only occurred recently. 
A very recent experiment~\cite{An2018} reports the observation of an SPME predicted for the long-range hopping $t_1$-$t_2$ bichromatic 1D incommensurate lattice~\cite{SPME07,SPME08} in a rather small $21$-site system.
This is particularly germane in the context of the fact that generic 3D disordered systems do indeed have SPMEs, and hence studying SPMEs in controlled 1D optical lattice experiments is important. 
Given the limited number of experimental realizations in spite of a great deal of theoretical activity, it is crucial to continue to identify and study other experimentally feasible 1D models with an SPME. 
This is the purpose of the current theoretical study where two suitable systems are identified with SPME which should manifest robust intermediate phases as the system parameters are tuned.

In this work, we discuss two realistic 1D models where an SPME exists for a wide range of parameters. 
The first model is the Soukoulis-Economou model containing two incommensurate potentials in a 1D tight-binding lattice, which was proposed  in 1982~\cite{Soukoulis_Economou}. 
It is well established that this model contains an SPME. 
However, it has never been studied in the context of the single-particle intermediate phase. 
Here we demonstrate that this model in fact provides an ideal and realistic platform to study the coupling of extended and localized degrees of freedom, because it contains a much wider single-particle intermediate phase than the one realized in recent experiments~\cite{LuschenSPME17}, yet does not require a fundamentally different experimental setup. 
In addition, we will discuss the critical properties of the eigenstates near the mobility edge in this model, emphasizing that this model is not a mere mixture of localized and extended single-particle orbitals such as the $s$-$p$ model discussed in Ref.~\cite{Xiao17}. 
Specifically, it is known that when approaching the mobility edge $E_c$ from the side of localized states, the localization length of the eigenstates will keep increasing and eventually diverge at $E_c$. 
As a result, the inverse localization length, or the Lyapunov exponent $\gamma$, has a well-defined scaling behavior near $E_c$. 
We thus study the scaling exponent of the Lyapunov exponent of this model, and compare it to that of a related family of incommensurate lattice models. 
We find that unlike 3D Anderson models, where this scaling exponent seems to be independent of the disorder realizations~\cite{Slevin2014}, for 1D incommensurate lattice models the scaling exponent strongly depends on the model details. 
Thus, the concept of a universal localization exponent does not exist in 1D quasiperiodic systems with the critical exponent being crucially dependent on the Hamiltonian parameters.

The second experimentally relevant model we consider is a quasi-1D system consisting of two parallel 1D {\AAM} chains of equal size. 
Specifically, we keep one chain completely free of disorder (i.e., no  incommensurate potential) while have the other subject to a quasiperiodic potential with strength $V_2$. 
Thus, one chain by itself is simply a free 1D tight binding lattice with only extended states whereas the other chain by itself has the usual {\AAM} self-dual transition with all states being localized or extended depending on the strength $V_2$ of the incommensurate potential. 
In addition, we introduce a nonzero (and uniform) inter-chain hopping term $\tp$ along each rung of the ladder.
The presence of the inter-chain hopping breaks the self-duality and leads to nontrivial effects as shown in our work.  (One could think of the first chain with all extended states acting as a `bath' for the second chain with the incommensurate potential, but the coupling between them is simply a hopping term with no interaction in the problem.) 
Our main findings for this model are twofold. 
First, we have identified that the system contains a wide single-particle intermediate phase, especially for small $\tp/t$ (where $t$ is the nearest-neighbor hopping term within each chain, taken to be the same for both chains). 
Therefore, this is another promising system to explore the effects of an SPME. 
(Since the single {\AAM} chain has already been studied extensively experimentally~\cite{Roati2008,Schreiber_2015} in the ultracold atom optical lattice systems, we believe that our two-coupled chain system should be straightforward to implement in the laboratory.) 
Second, our results reveal the complexity of interchain coupling between localized and extended degrees of freedom, even without interactions. 
Specifically, assuming a sufficiently large incommensurate potential strength $V_2 > 2t$ (so that all states are localized in the isolated second chain in the absence of interchain hopping), the effect of a small $\tp/t$ is to delocalize the originally localized orbitals at $\tp=0$. 
By contrast, when $\tp/t$ is large (while keeping $V_2 > 2t$), the effect of the inter-chain coupling is to destroy the originally extended orbitals at $\tp=0$. 
The above results are summarized in a rich localization phase diagram, which is quite nontrivial. 

The main goal of this work is to motivate further experimental studies of systems manifesting wide and robust intermediate phases in the single-particle limit. 
Once such noninteracting intermediate phases are experimentally established, one can further study their stability in the presence of finite interaction to investigate the deep question of whether the SPME and, consequently, the intermediate phase survives inter-particle interaction. 
The two systems we propose here should be essentially ideal systems for studying the interplay of localized and extended states in the context of MBL to answer the important question of whether the presence of a small bath of extended states immediately destroys MBL or whether such a destruction must always go through an intermediate nonergodic metallic phase lying between ETH and MBL phases. 
The fact that the intermediate phase in the two models we study is stable over a large parameter range makes them particularly attractive for future MBL experimental investigations. 

The structure of this paper is the following. 
In Section~\ref{Section:SEModel} we present our results on the {\SE} model, including the scaling analysis of the Lyapunov exponents near the mobility edge $E_c$. 
In Section~\ref{Section:CoupledChain} we present our results on the coupled {\AAM} chain model and discuss its localization phase diagram. 
In Section~\ref{Section:Discussion} we draw a careful distinction between the existence of a single-particle mobility edge (SPME) and the existence of a single-particle intermediate phase (SPIP) in a model, and discuss its implications when we generalize it to a many-body system. 
Section~\ref{Section:Summary} summarizes the main results of this work and presents some concluding remarks. 
In Appendix~\ref{Appendix:A} some additional properties of the  single-particle eigenstates in these two models are presented. 
Finally, in Appendix~\ref{Appendix:B}, we establish certain duality properties among various 1D incommensurate models showing that various classes of quasiperiodic models exhibiting SPME are connected by duality transformations.

\section{The Soukoulis-Economou model \label{Section:SEModel}}
The Soukoulis-Economou model~\cite{Soukoulis_Economou} was one of the first proposals of one-dimensional quasiperiodic models with an SPME. 
It is a tight-binding model with nearest-neighbor hopping terms as well as two quasiperiodic on-site potentials, 
\begin{align}
	E u_n = t(u_{n-1} + u_{n+1}) + V_0\left[\cos(Qn) + V_1 \cos(2Qn)\right] u_n. \label{Eq:Soukoulis-Economou}
\end{align}
In the above equation, $t$ is the nearest-neighbor hopping, $V_0$ and $V_1$ determine the strength of the two quasiperiodic potentials, and $Q$ is the wavevector of the potential. In addition, $u_n$ is the wave function amplitude on site $n$, and $E$ is the energy eigenvalue. 
With $V_1=0$, this model is simply the {\AAM} model~\cite{aubry1980analyticity} where all states are extended ($V_0<2t$) or localized ($V_0>2t$) without any SPME or SPIP---all states are either localized or extended, depending the dimensionless parameter $V_0/t$. 
For finite $V_1$, however, this model manifests a mobility edge in the single-particle energy spectrum. (Some details of our numerical results are given in Appendix~\ref{Appendix:A}.)

We study the localization properties of this model for the following reasons. 
First, it is relatively easy to implement this model in laboratory optical lattices, as the on-site incommensurate potential can be created by using two lasers with commensurate frequencies. 
As a result, its laboratory implementation only requires a slight modification of the existing experimental setup~\cite{LuschenSPME17}. 
Second, as we show in this section, the {\SE} model generally manifests a much wider (as a function of system parameters)  single-particle intermediate phase than the one realized in the recent experiment~\cite{LuschenSPME17}. 
From the experimental perspective, it is often desirable to have a broad single-particle intermediate phase because it allows for a wide range of tunability, which in turn makes it easier to detect the possible existence of a many-body intermediate phase when the interaction is turned on. 
Therefore, this model provides a very attractive platform to explore the physics of many-body localization in a system with an SPME. 
These considerations motivate us to study the localization properties of this model in details, hoping that our results can draw the community's attention to this very interesting model for laboratory measurements.

\subsection{Single-particle intermediate phase in the Soukoulis-Economou model}

\begin{figure}[!]
\includegraphics[width=3.0in]{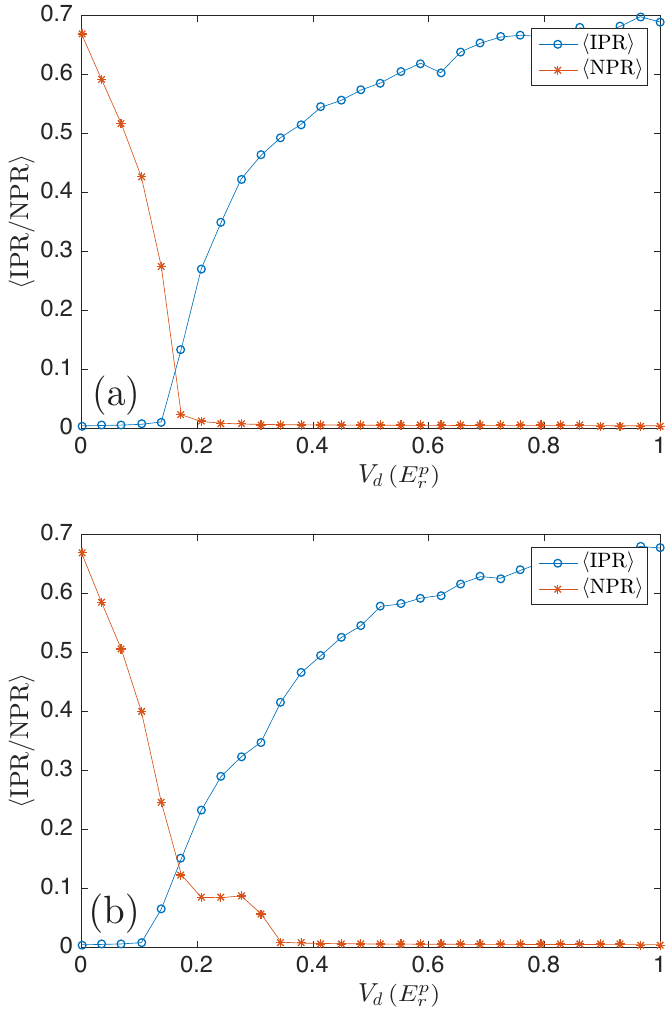}
\caption{\label{Fig:IPR_Cont} The width of the single-particle intermediate phase in (a) the bichromatic potential model and (b) the Soukoulis-Economou model. Here the primary lattice has a depth of $V_s = 8E_r^p$, the incommensurate ratio is $\alpha = 532/738$, and the system size is $L = 369$. 
Moreover, we set $V_d'/V_d = 1/3$ in the Soukoulis-Economou model.}
\end{figure}

In order to make a proper comparison with the bichromatic potential model realized in the recent experiment~\cite{LuschenSPME17}, we start from the continuum limit of the {\SE} model. 
To start, consider the following single-particle continuum Hamiltonian, 
\begin{align}
	H_\text{continuum} = -\dfrac{\hbar^2}{2m}\dfrac{d^2}{dx^2} + V(x), \label{Eq:ContinuumModel}
\end{align}
where $m$ is the mass of the particles, and $V(x)$ is given by the following `trichromatic' potential 
\begin{align}
	V(x) = \dfrac{V_p}{2}\cos(2k_px) + \dfrac{V_d}{2}\cos(2k_dx+\phi) + \dfrac{V_{d}'}{2}\cos(4k_dx + \phi). \notag
\end{align}
In the above model, $k_p$ and $k_d$ are the wavevector of the primary and detuning lattice, respectively, and we define the incommensurate ratio $\alpha = k_d/k_p$. 
In addition, $V_p$, $V_d$, and $V_d'$ represent the depth of the respective lattice. 
For convenience, we also introduce the recoil energy of the primary lattice $E_r^p= \hbar^2k_p^2/(2m)$ as the energy unit for this continuum model. 
When the third term in $V(x)$ is switched off ($V_{d}'=0$), so that the trichromatic potential becomes a bichromatic potential, we have exactly the Hamiltonian that describes the experimental system in Ref.~\cite{LuschenSPME17}, which we will refer to as the \emph{bichromatic lattice model}. 
When it is switched on, however, the entire trichromatic Hamiltonian provides a continuum realization of the Soukoulis-Economou model given in Eq.~\eqref{Eq:Soukoulis-Economou}. 
Such a continuum Hamiltonian can be solved by discretizing the real-space coordinates and then diagonalizing the resulting matrix. 
In addition, only states in the lowest band will be retained in our calculation, as they are the relevant states for the tight-binding Hamiltonian in Eq.~\eqref{Eq:Soukoulis-Economou}. 

The presence of a single-particle intermediate phase (where both localized and extended eigenstates exist in the energy spectrum) can be demonstrated by simultaneously calculating the average (over all eigenstates) inverse participation ratio (IPR) and the average normalized participation ratio (NPR) of the single-particle eigenstates~\cite{Xiao17}. 
For the $i$th eigenstate $u_n^{(i)}$, they are defined as 
\begin{align}
	\text{IPR}^{(i)} = \sum_{n} \lvert u_n^{(i)}\rvert ^4, \quad 
	\text{NPR}^{(i)} = \left[L\sum_{n} \lvert u_n^{(i)}\rvert ^4\right]^{-1}, 
\end{align}
where $L$ is the size of the system, and the sum is over the lattice sites denoted by $n$. 
We further average these two quantities over all eigenstates (in the lowest energy band), and define the single-particle intermediate phase as the regimes in which both $\avgIPR$ and $\avgNPR$ are finite~\cite{Xiao17} (where $\average{\cdot}$ indicates an average over all eigenstates), also see Table~\ref{Table:SPIP}. 
Figure~\ref{Fig:IPR_Cont} compares the single-particle intermediate phase in the bichromatic lattice model and the Soukoulis-Economou trichromatic model when the primary lattice has a depth of $V_p = 8E_r^p$. 
Note that in this figure we only keep a small system size $L=369$, in order to simulate current experimental capabilities~\cite{LuschenSPME17}. 
We find that at this relatively deep primary lattice limit, the intermediate phase in the bichromatic lattice model is negligible, and the system can be well approximated~\cite{Xiao17,LuschenSPME17} by the Aubry-Andre model~\cite{aubry1980analyticity}
\begin{align}
	E u_n = t(u_{n-1} + u_{n+1}) + V \cos(2\pi\alpha n + \phi). 
\end{align}
In contrast, even for such a deep primary lattice potential, the Soukoulis-Economou model already carries a fairly wide intermediate phase, which makes it a promising experimental realization of a single-particle intermediate phase. 
From the theoretical perspective, it is also preferable to have a wide intermediate phase when the primary lattice is deep. 
The reason is that in such a limit the tight-binding model in Eq.~\eqref{Eq:Soukoulis-Economou} will already be an accurate description of the experimental system, and thus there is no need to invoke the continuum description. 
As a result, we will only work with the tight-binding model [Eq.\eqref{Eq:Soukoulis-Economou}] in our subsequent discussions because we assume the experimentally easily accessible deep primary lattice potential limit.

\subsection{Lyapunov exponent in the Soukoulis-Economou model}

The intermediate phase in the {\SE} model (or in any incommensurate lattice model with an SPME) is more than a simple combination of extended and localized orbitals in the same system. 
In fact, the latter scenario can be generated by simply considering a mixture of two species of particles, one localized while the other extended~\cite{Xiao17}. 
The crucial feature of this family of incommensurate lattice models is the existence of the SPME itself. 
In particular, as one approaches the SPME $E_c$ from the side of localized eigenstates, the localization length $\xi$ of the eigenstates will diverge near $E_c$. 
As a result, the Lyapunov exponent $\gamma \equiv \xi^{-1}$~\cite{SPME04,SPME06,Kramer_MacKinnon_1993,AndersonTransitions_RMP_2008} satisfies the following scaling behavior near $E_c$, 
\begin{align}
	\gamma(E) \sim \abs{E-E_c}^\delta.  \label{Eq:ScalingExponent}
\end{align} 
Therefore, an extended (localized) eigenstate is characterized by a vanishing (finite) Lyapunov exponent. 
{By definition, the localization length is infinite (or $\gamma=0$) on the extended side. } 

To illustrate this point, here we calculate the Lyapunov exponent $\gamma$ and the corresponding scaling exponent $\delta$ in several incommensurate lattice models. 
We first explain the method we use and then present the numerical results. 

\subsubsection{Calculation of the Lyapunov exponent}
In general, when the tight-binding model involves nearest-neighbor hopping terms only, i.e.,  
\begin{align}
	E u_n = t(u_{n-1} + u_{n+1}) + V_n u_n, \label{Eq:TBModel}
\end{align}
where $V_n$ is the on-site potential term, 
the Lyapunov exponent can be calculated conveniently by the formula~\cite{SPME06}
\begin{align}
	\gamma(E_i) = \dfrac{1}{N-1} \sum_{j\neq i} \ln\abs{\dfrac{E_i-E_j}{t}}, 
\end{align}
where $N$ is the number of lattice sites, and $E_i$ are the energy eigenvalues. 
More generally, however, the Lyapunov exponent can be evaluated using the transfer matrix technique as we now explain. 
Specifically, the tight-binding model in Eq.~\eqref{Eq:TBModel} can be cast into the following form, 
\begin{align}
	\begin{pmatrix} u_{n+1} \\ u_n \end{pmatrix} = 
	T_n 
	\begin{pmatrix} u_{n} \\ u_{n-1} \end{pmatrix}, 
\end{align}
where the transfer matrix $T_n$ is given by 
\begin{align}
	T_n = \begin{pmatrix} E - V_n & -t \\ t & 0 \end{pmatrix}. 
\end{align}
As a result, we can define a new matrix 
\begin{align}
	\bm{\Lambda} = \lim_{N\to +\infty} \left(\bm{T}_N^\dagger \bm{T}_N\right)^{1/(2N)}, \quad \bm{T}_N \equiv \prod_{k=1}^{N} T_k. \label{Eq:Lambda}
\end{align}
The fact that such a matrix limit exists is guaranteed by Oseledec's ergodic theorem~\cite{Oseledec_Theorem}. 
The Lyapunov exponents are then given by $\lambda_i = \ln \Lambda_i$, where $\Lambda_i\in \mathbb{R}$ are the eigenvalues of the matrix $\bm{\Lambda}$. 
For the present case, the analysis is straightforward because $\bm{\Lambda}$ is a $2\times2$ symplectic matrix. Therefore, its two eigenvalues are equal in magnitude and opposite in sign, and the Lyapunov exponent can be taken to be the magnitude of either eigenvalue. 

\begin{figure}[!]
\includegraphics[width = 3.3in]{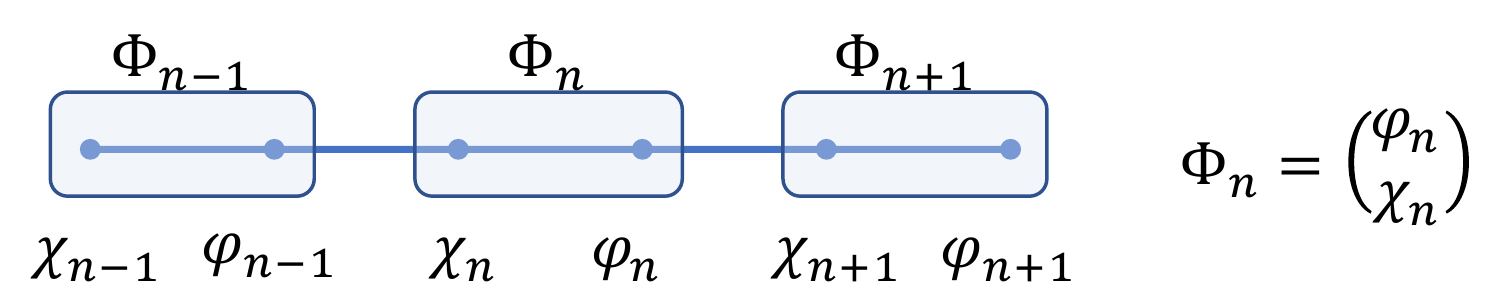}
\caption{\label{Fig:TransferMatrix} 
Construction of supercells in the $t_1$-$t_2$ model in Eq.~\eqref{Eq:t1t2Model}. 
Although the original model has long-range hopping terms, the supercells only have nearest-neighbor hopping terms between them. }
\end{figure}

The above transfer matrix technique is very versatile, and can be easily generalized to tight-binding models with long-range hopping terms~\cite{Chua2016}. 
As an example, we consider the $t_1$-$t_2$ model~\cite{SPME07,SPME08}, 
\begin{align}
	Eu_n = t_1(u_{n-1} + u_{n+1}) + t_2(u_{n-2} + u_{n+2}) + V_n u_n, \label{Eq:t1t2Model}
\end{align}
which is a tight-binding model with both nearest-neighbor and next-nearest-neighbor hopping terms. 
In order to apply the transfer matrix technique to models with long-range hopping terms, we group the lattice sites into supercells~\cite{Chua2016} so that the hopping between supercells is limited to the nearest neighbors only (see Fig.~\ref{Fig:TransferMatrix} for an illustration). 
To be specific, we adopt the following convention for the wave function amplitudes, 
\begin{align}
	\chi_n \equiv u_{2n-1}, \; \varphi_n \equiv u_{2n}, \; n = 1, 2, \cdots, 
\end{align}
and introduce a supercell wave function $\Phi_n^\top = \begin{pmatrix} \varphi_n & \chi_n \end{pmatrix}$. 
As a result, the $t_1$-$t_2$ model can be written in terms of $\Phi_n$ as follows, 
\begin{align}
	E\Phi_n = J \Phi_{n+1} + M_n \Phi_n + J^\dagger \Phi_{n-1}, \label{Eq:TransferMatrix2D}
\end{align}
where the matrices $J$ and $M_n$ are given by 
\begin{align}
	J = \TwoDMatrix{t_2}{1}{0}{t_2}, \; 
	M_n = \TwoDMatrix{V_{2n}}{t_1}{t_1}{V_{2n-1}}. 
\end{align}
Now observe that for a nonsingular $J$ (which is the case in the $t_1$-$t_2$ model), the above equation can be rewritten conveniently as follows~\cite{Chua2016}, 
\begin{align}
	\begin{pmatrix} \Phi_{n+1} \\ \Phi_{n} \end{pmatrix}
	&= 
	\TwoDMatrix{J^{-1}(E\mathds{1}-M)}{-J^{-1}J^\dagger}{\mathds{1}}{0} 
	\begin{pmatrix} \Phi_{n} \\ \Phi_{n-1} \end{pmatrix} \notag\\
	&= T_n \begin{pmatrix} \Phi_{n} \\ \Phi_{n-1} \end{pmatrix},  
\end{align}
The above analysis resulted in a $4\times4$ transfer matrix $T_n$ for the {\ttM}. 
The Lyapunov exponent can be subsequently obtained by inserting this transfer matrix into Eq.~\eqref{Eq:Lambda}. 
Note that in this case the resulting $\bm{\Lambda}$ is a $4\times4$ symplectic matrix, whose four eigenvalues appear in pairs and can be denoted by $\pm \lambda_<$ and $\pm\lambda_>$, respectively, with $\lambda_> > \lambda_< > 0$. 
The Lyapunov exponent will then be determined by $\lambda_<$, because it is related to the longest localization length in the system. 

\subsubsection{Lyapunov exponents in several incommensurate lattice models}

Having set up the formalism, we now use it to calculate the Lyapunov exponents of several incommensurate lattice models. 
Specifically, in addition to the {\SE} model, we will consider three other models. 
The first one is a generalized {\AAM} (GAA) model. 
It has the same form as Eq.~\eqref{Eq:TBModel}, with the on-site potential term given by~\cite{GAAModel_Ganeshan_2015}
\begin{align}
	V_n^{\text{(GAA)}} = 2\lambda \dfrac{1-\cos(2\pi\alpha n + \phi)}{1+\beta\cos(2\pi\alpha n + \phi)}. \label{Eq:GAA}
\end{align}
The second is a cosine potential model, again given by Eq.~\eqref{Eq:TBModel}, with the following on-site potential term~\cite{SPME06}
\begin{align}
	V_n^{\text{(cosine)}} = \lambda \cos(\pi\alpha n^\nu). \label{Eq:Cosine}
\end{align}
Finally, we will consider the $t_1$-$t_2$ model in the form of Eq.~\eqref{Eq:t1t2Model}, with the potential term given by $V_n = \cos(2\pi\alpha n + \phi)$. 

\begin{figure}[!]
\includegraphics[width = 3.0in]{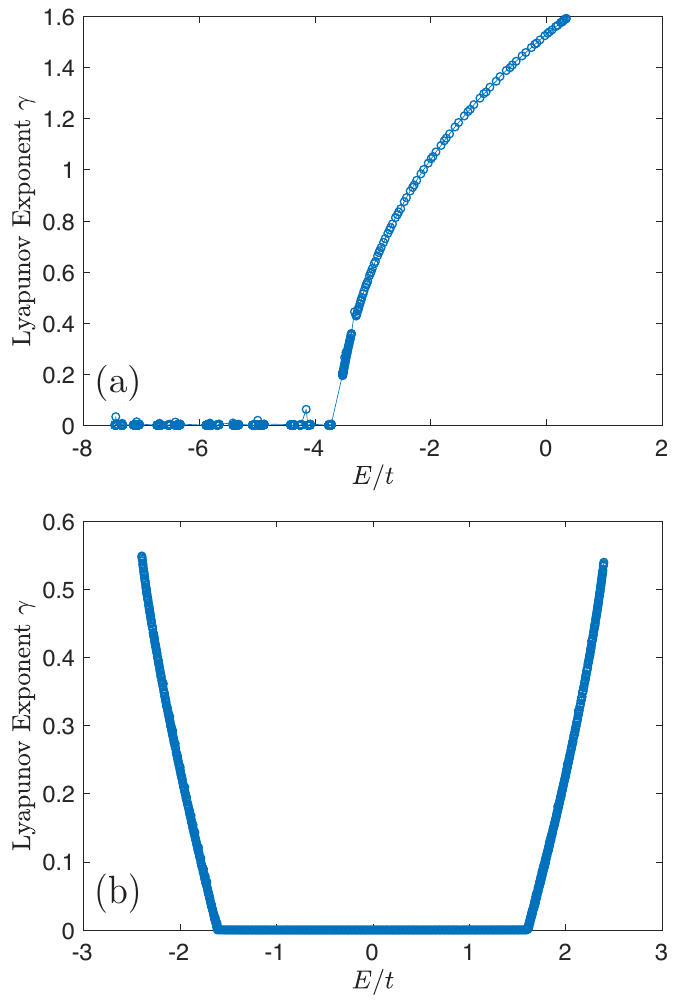}
\caption{\label{Fig:Lyapunov_GAA_Cosine} Lyapunov exponent in the (a) generalized {\AAM} (GAA) model and (b) the cosine potential model. The parameters for the GAA model are given by $\alpha = 2/(\sqrt{5}-1)$, $\beta = -0.99$, $\lambda = -2.80$. The parameters for the cosine potential model are given by $\lambda = 0.4$, $\nu = 0.7$, $\pi\alpha = 0.2$. The system size in both cases is $L = 10^4$.}
\end{figure}

Figure~\ref{Fig:Lyapunov_GAA_Cosine} shows the calculated Lyapunov exponent in the GAA model and the cosine potential model. 
We can see that both models have a very nice scaling behavior near $E_c$, as the Lyapunov exponents decay smoothly towards $E_c$ from the localized side of the spectrum. 
In addition, we find that the corresponding scaling exponent [cf. Eq.~\eqref{Eq:ScalingExponent}] is $\delta \simeq 0.53$ for the GAA model, while $\delta \simeq 1.13$ for the cosine potential model.

\begin{figure}[!]
\includegraphics[width=3.0in]{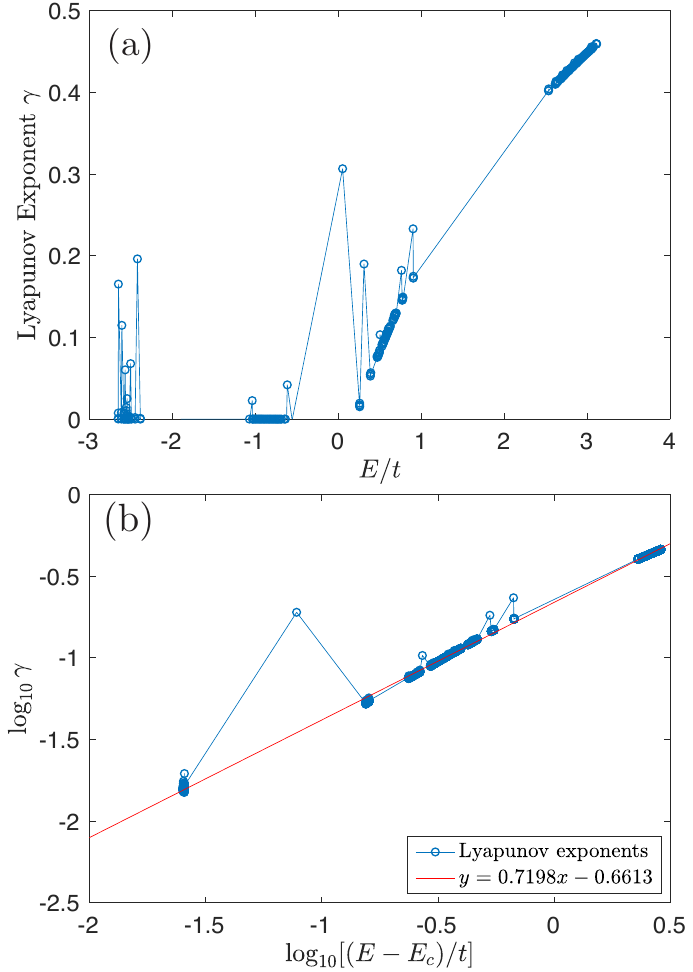}
\caption{\label{Fig:SEModel} (a) Lyapunov exponent $\gamma$ and (b) scaling analysis in the Soukoulis-Economou model. The corresponding model parameters are given by $Q = 2/(\sqrt{5}-1)$, $V_0 = 1.90$, and $V_1 = 1/3$. In addition, the system size is $L = 10^4$. }
\end{figure}

Figure~\ref{Fig:SEModel} shows the calcualted Lyapunov exponent for the {\SE} model. 
One can see immediately that the spectrum again contains both localized and extended eigenstates, and that the SPME is located near $E_c = -0.2t$. 
However, the Lyapunov exponent in this model does not seem to follow an apparent clean scaling behavior near the SPME that was seen in the previous two models. 
Nonetheless, we are still able to extract a scaling exponent for the Lyapunov exponent by only retaining states very close to $E_c$, as shown in Fig.~\ref{Fig:SEModel}(b). In particular, we determine that $\delta \simeq 0.7198$ for this model. 
Apparently, the scaling regime for the Soukoulis-Economou model is rather small, making the extraction of a localization exponent somewhat of a challenge compared to the GAA model.

\begin{figure}[!]
\includegraphics[width=3.0in]{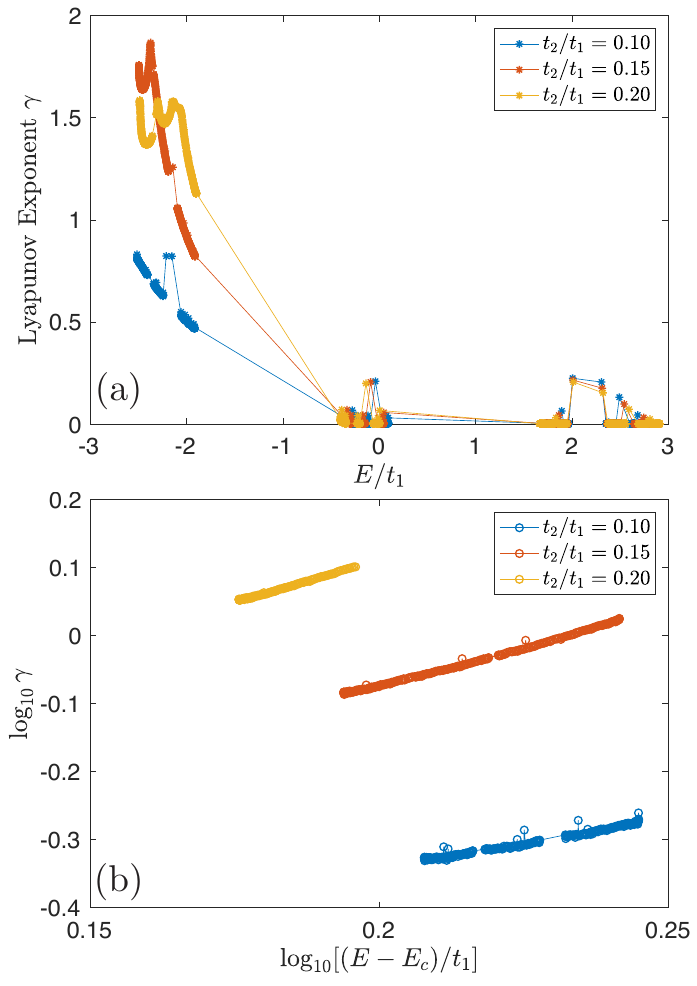}
\caption{\label{Fig:t1t2Model} (a) Lyapunov exponent $\gamma$ and (b) scaling analysis in the {\ttM}. Here we set $t_1 = 1$ as the energy unit, $\alpha = 2/(\sqrt{5}-1)$, and the system size is $L=10^4$. In particular, the scaling exponents for the three cases are the following: $\delta\simeq 1.4790$ for $t_2/t_1=0.10$, $\delta\simeq 2.2540$ for $t_2/t_1=0.15$, and $\delta\simeq 2.442$ for $t_2/t_1=0.20$. }
\end{figure}

Finally, we analyze the Lyapunov exponent in the {\ttM}. As shown in Fig.~\ref{Fig:t1t2Model}, the scaling behavior of the Lyapunov exponent near the SPME is again not apparent in this model. 
Following the same method as Fig.~\ref{Fig:SEModel}, we are able to extract the scaling exponent $\delta$ for three different values of $t_2/t_1$ [see Fig.~\ref{Fig:t1t2Model}(b)]. 
It can be seen that the exponent $\delta$ is very sensitive to the $t_2/t_1$ ratio. 
Again, the scaling regime appears to be small for the $t_1$-$t_2$ model similar to the Soukoulis-Economou model.  
In addition, the dependence on $t_2/t_1$ indicates a lack of universality of the localization exponent (or in other words, each value of $t_2/t_1$ represents a universality class by itself). 

One may suspect that the $V_0$-$V_1$ model of Soukoulis-Economou~\cite{Soukoulis_Economou} has some connection to the $t_1$-$t_2$ model of Biddle et al.~\cite{SPME08} since both of them have two dimensionless parameters breaking the duality symmetry of the AA model, albeit an extra incommensurate potential ($V_1$) for Ref.~\cite{Soukoulis_Economou} and an extra hopping term ($t_2$) for Ref.~\cite{SPME08} compared with the AA model.  
In addition, our calculation of the Lyapunov exponent also demonstrates a similarity between them in that both models have very narrow critical regimes (i.e., $E$ very close to $E_c$) with very nonuniversal exponents.  
In Appendix~\ref{Appendix:B} we establish a hitherto unknown connection between these two incommensurate models (both with SPME) by showing that the Biddle et al. $t_1$-$t_2$ next-nearest neighbor hopping model~\cite{SPME08} is in fact dual to the Soukoulis-Economou $V_0$-$V_1$ trichromatic potential model~\cite{Soukoulis_Economou}. 
In addition, we also theoretically obtain in Appendix~\ref{Appendix:B} the dual model for the long-ranged hopping model of Ref.~\cite{SPME07}, which was solved exactly showing the existence of an SPME. 

The results in this section clearly demonstrate the non-universality of the localization transition near the mobility edge in this family of 1D incommensurate lattice models. 
Such transitions depend sensitively on all the details of how the duality of the AA model is broken. 
This is different from the 3D Anderson model where the corresponding mobility edge exponent is thought to be universal, independent of the details of the random disorder. 
In particular, the correlation length exponent obtained from the finite size scaling analysis of the Lyapunov exponent in the 3D Anderson model seems to be $\sim1.57$ for different models of disorder~\cite{Slevin2014}.

\section{Two coupled 1D Aubry-Andre chains \label{Section:CoupledChain}}

\begin{figure}[!]
\includegraphics[width = 3.3in]{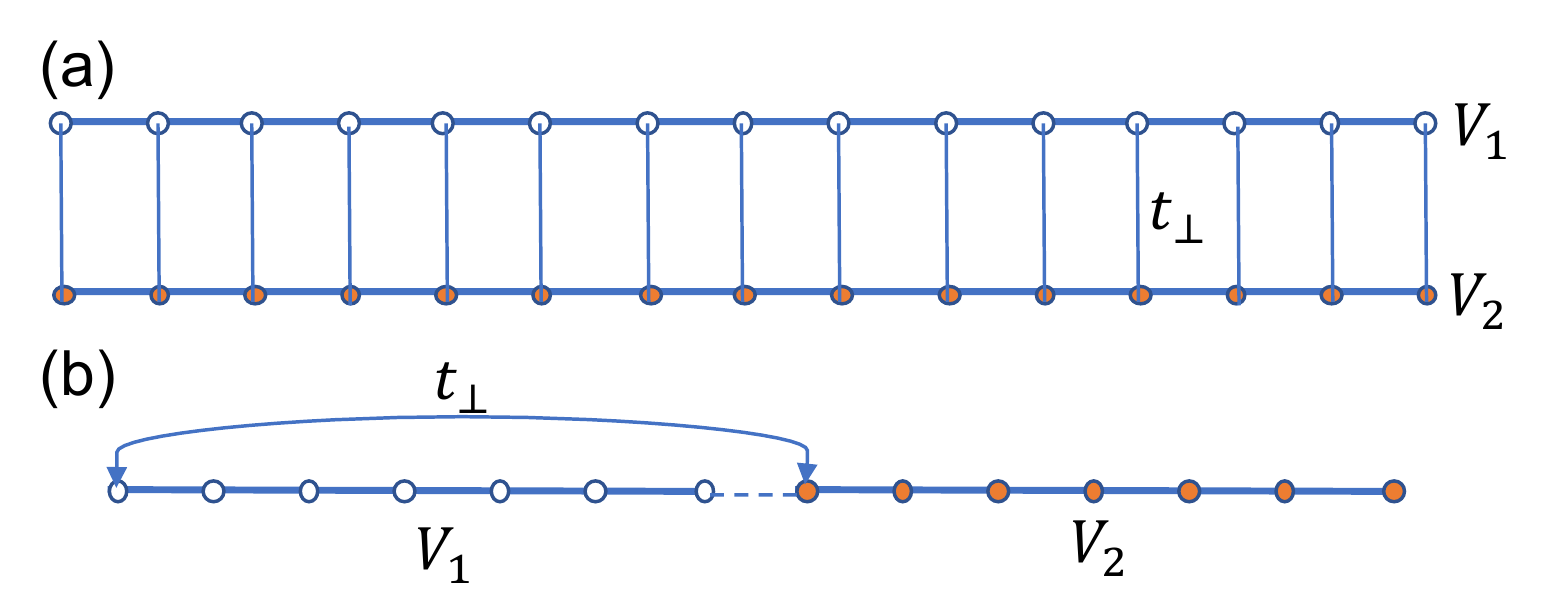}
\caption{\label{Fig:CoupleChainModel} (a) Illustration of the coupled chain model in Eq.~\eqref{Eq:CoupledChain}. The two chains are subject to a quasiperiodic potential of strength $V_1$ and $V_2$, respectively, and there is an inter-chain tunneling $\tp$ that couples them. (b) An equivalent formulation of the above problem in a one-dimensional system. }
\end{figure}

We now turn to a very different model in which an SPME can be found. 
As mentioned already, our reason for studying this model theoretically is the easy laboratory accessibility of the corresponding experimental system in cold atom optical lattices. 
In particular, we consider two coupled 1D {\AAM} chains with different quasiperiodic potential strengths $V_1$ and $V_2$, as shown in Fig.~\ref{Fig:CoupleChainModel}(a). 
The corresponding Hamiltonian is given by $H = H_1 + H_2 + H_{\perp}$, where 
\begin{align}
	H_{i=1,2} &= t\sum_{\langle nm\rangle} \left(c^{\dagger}_{i, n}c_{i, m}+\text{h.c.}\right) + V_i \sum_{n} \cos(2\pi\alpha n + \phi)c^{\dagger}_{i,n}c_{i,n}, \notag\\
	H_{\perp} &= \tp \sum_{n} \left(c^{\dagger}_{1,n}c_{2,n} +\text{h.c.} \right). \label{Eq:CoupledChain}
\end{align}
We keep one chain completely clean $(V_1 = 0)$ while the other subjected to a quasiperiodic potential with strength $V_2$, and study how the inter-chain tunneling $\tp$ affects the localization properties of the coupled system. 
In this section we will keep $\alpha = 2/(\sqrt{5}-1)$ in all our calculations. 
For $\tp=0$, this is a completely understood problem. 

It is interesting to note that this coupled chain model is equivalent to a single one-dimensional chain with long-range hopping terms, as shown in Fig.~\ref{Fig:CoupleChainModel}(b). 
Specifically, consider a system with $2L$ sites and imagine that the first $L$ sites are subject to a quasiperiodic potential with strength $V_1$, while the other $L$ sites are subject to a quasiperiodic potential with strength $V_2$. 
Now we introduce a long-range hopping $\tp$ between the site $i$ ($1\leq i\leq L$) and the site $i+L$. 
The Hamiltonian of such a 1D system is exactly the same as that of the coupled chain model in Fig.~\ref{Fig:CoupleChainModel}(a). 
Our discussion below will be based on the coupled chain model in Eq.~\eqref{Eq:CoupledChain}, but they are applicable for both cases. 
Meanwhile, the breaking of the {\AAM} duality is manifest in the single 1D model with long-range hopping terms, and hence we anticipate an SPME in the system. 

\subsection{Localization transition at a fixed \texorpdfstring{$V_2$}{V2}}

\begin{figure}[!]
\includegraphics[width = 3.3in]{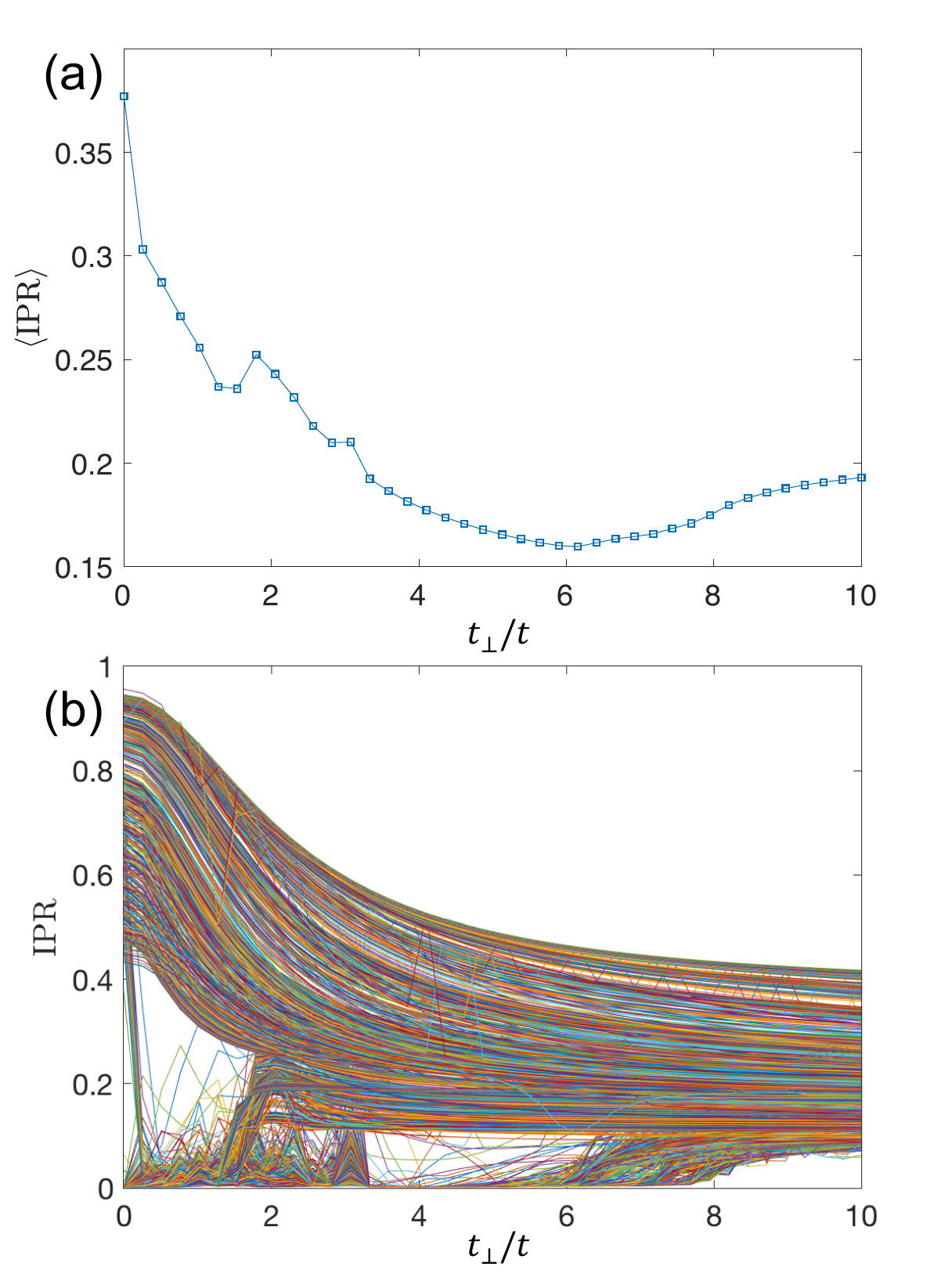}
\caption{\label{Fig:CoupledChain1a} (a) Averaged IPR in the coupled chain model with $V_1 = 0$ and $V_2 = 5t$. (b) IPR of each individual eigenstate in the same model. Here the size of each chain is $L=1000$, and $\alpha = 2/(\sqrt{5}-1)$.} 
\end{figure}

We first fix the value of $V_2$ in the localized phase ($V_2>2t$) and discuss how the localization property of the system varies as a function of $\tp$. 
The $\tp = 0$ limit is simple, as half of the eigenstates are localized while the other half are extended, and the entire spectrum is simply a mixture with no SPME. 
In a system of $L=1000$, the average IPR over all eigenstates is $\avgIPR \simeq 0.4429$. 
Note that this case is a trivial SPIP since the system is, by construction, a mixture of extended and localized states. 

When $\tp > 0$ the two chains are coupled, and an interesting question we can ask is whether the system will become more extended or more localized as $\tp$ is increased. 
A naive expectation is that the clean chain will help delocalize the states on the disordered chain, and will thus always drive the coupled system towards an overall more extended state. 
However, such an expectation is only correct for a small $\tp$, and the localization property of the coupled system can be rather complicated for a large $\tp$. 
This conclusion can be illustrated by the $\avgIPR$ plot in Fig.~\ref{Fig:CoupledChain1a}, in which we take $V_1=0$ and $V_2 = 5t$. 
Specifically, we can see that for $0<\tp/t<1$ the originally localized states (at $\tp=0$) start to become more extended, while the originally extended states are not affected much. 
When $\tp/t>1$, however, the originally extended states (at $\tp=0$) start to get localized, and this trend continues until $\tp \sim 8t$, beyond which no purely extended states ($\text{IPR}\sim 0$) exist in the spectrum anymore.

\begin{figure}[!]
\includegraphics[width = 3.3in]{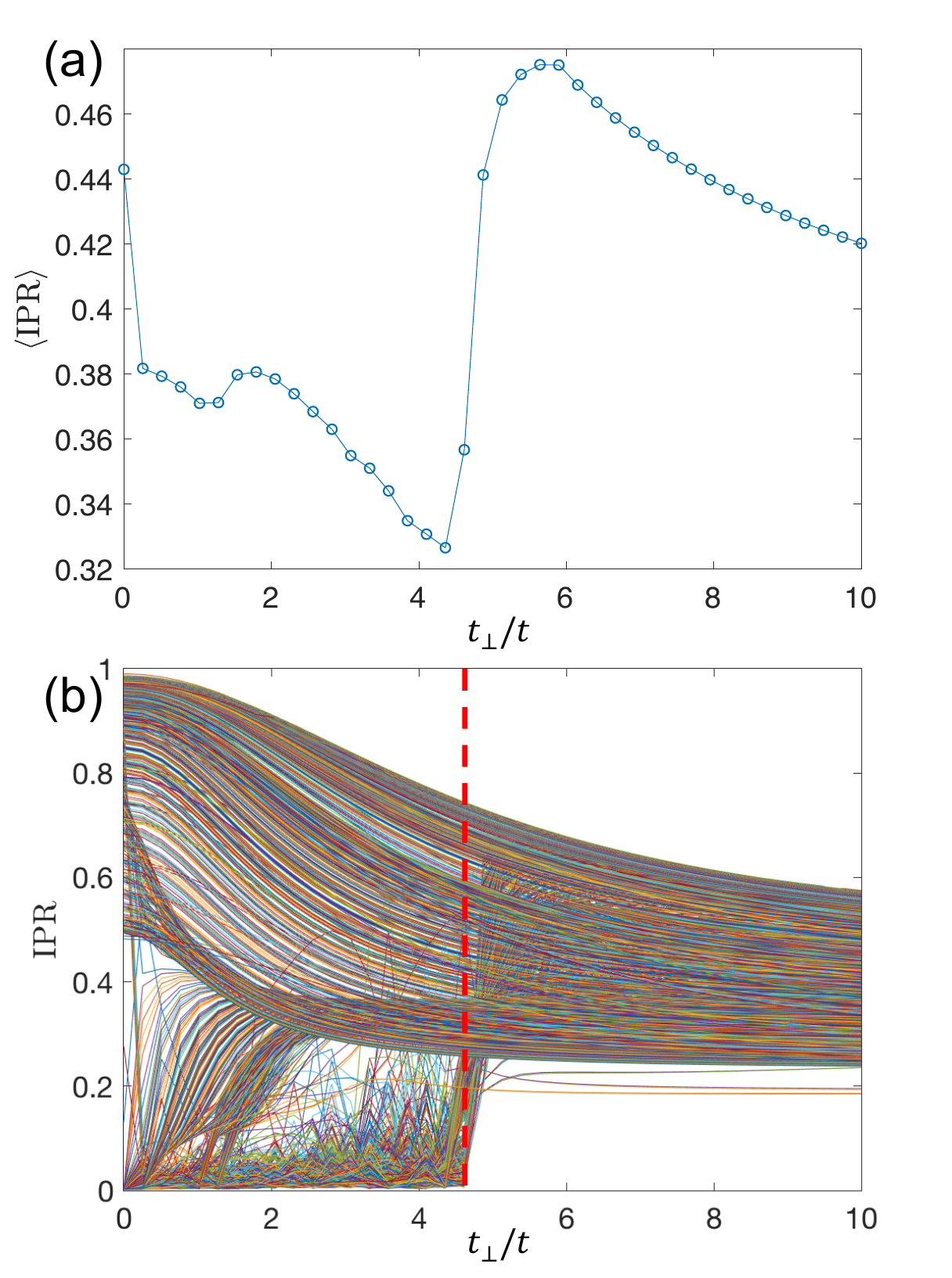}
\caption{\label{Fig:CoupledChain1b} (a) Averaged IPR in the coupled chain model with $V_1 = 0$ and $V_2 = 10t$. (b) IPR of each individual eigenstate in the same model. The dashed line marks the value of $\tp$ at which all extended states are destroyed. Here the size of each chain is $L=1000$, and $\alpha = 2/(\sqrt{5}-1)$.}
\end{figure}

In the above example we can see that the localization property of the coupled system (characterized by the averaged IPR) has a nontrivial and nonmonotonic dependence on $\tp$. 
In addition, for a modest $V_2$ value ($V_2\sim5t$), the averaged IPR at finite $\tp$ values is generally less than that at $\tp=0$, which is consistent with our expectation that a nonzero $\tp$ generally delocalizes the coupled system. 
However, even this expectation is no longer valid at large $V_2$ values. 
As an example, Fig.~\ref{Fig:CoupledChain1b} shows the localization properties of the coupled system when $V_2/t = 10$ (and $V_1=0$). 
It can be seen that for $5<\tp/t<6$ the averaged IPR clearly exceeds its value at $\tp = 0$. 
In addition, we note that the $\avgIPR$ plot has a curious cusp near $\tp/t \sim 4.5$. 
These two features can be understood by looking at the IPR plot for each individual eigenstate in Fig.~\ref{Fig:CoupledChain1b}(b). 
From the figure, it is clear that all extended states are destroyed at $\tp/t \sim 4.5$ (marked by the red dashed line), similar to what we observed near $\tp/t\sim 8$ for the $V_2/t = 5$ case in Fig.~\ref{Fig:CoupledChain1a}. 
However, in the present case the quasiperiodic potential is much stronger, causing the originally extended states to localize rather abruptly near $\tp/t \sim 4.5$, and thus leading to a very sharp rise of $\avgIPR$. 
It will be interesting to verify this nontrivial dependence of the spectrum on $\tp$ experimentally. 

\subsection{Localization transition at a fixed \texorpdfstring{$\tp$}{tp}}

\begin{figure}[!]
\includegraphics[width = 3.3in]{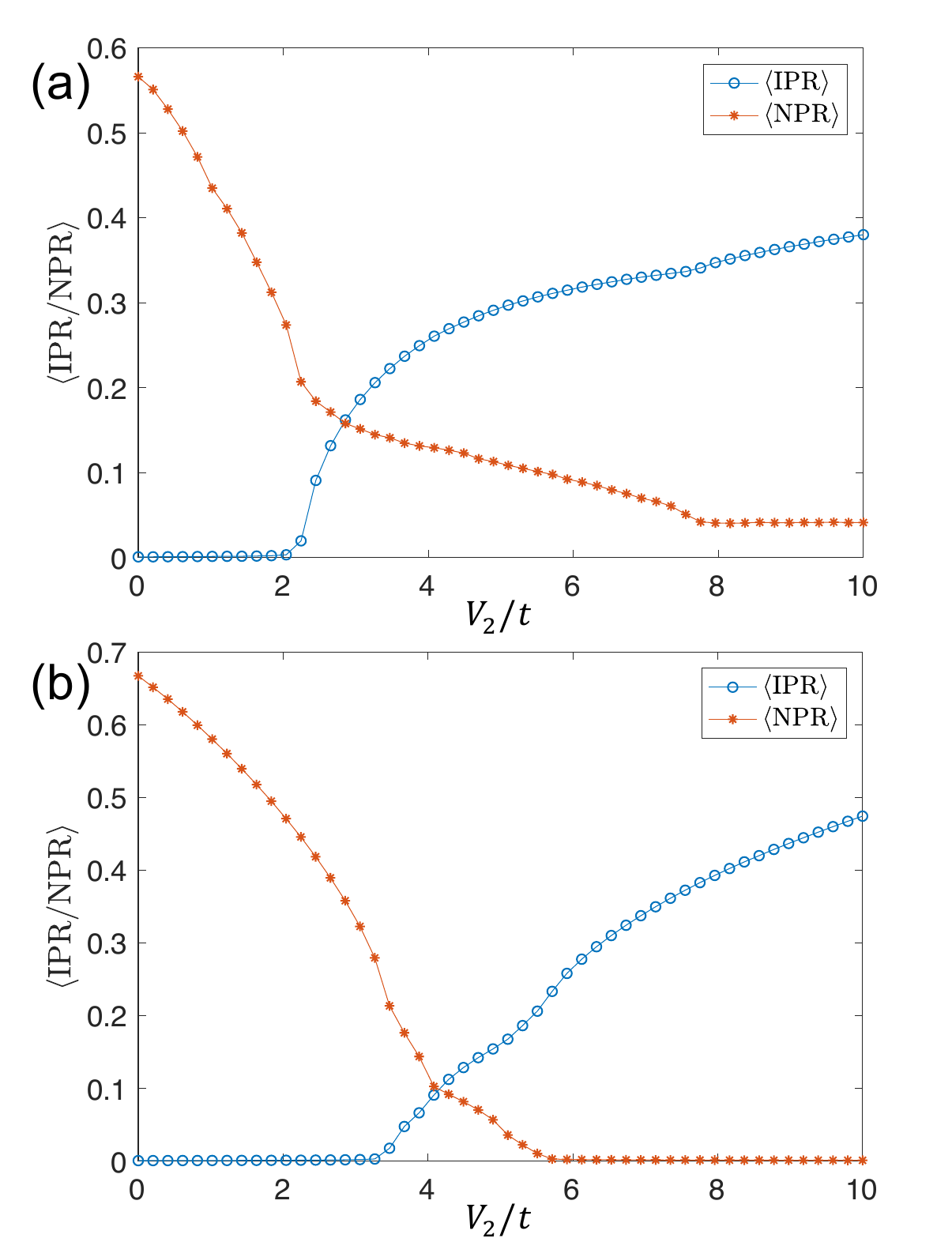}
\caption{\label{Fig:CoupledChain1c} Averaged IPR and NPR plot for the coupled chain model for (a) $\tp/t = 0.41$ and (b) $\tp/t = 5.92$. The size of each chain is $L=1000$, and $\alpha = 2/(\sqrt{5}-1)$.}
\end{figure}

Having studied how the inter-chain hopping $\tp$ affects the localization property of the coupled system, we now study the effect of varying $V_2$ with $\tp$ fixed (and keeping $V_1=0$). 
The $\tp=0$ limit is again easy to understand: both chains are extended when $V_2/t < 2$, whereas one chain will be completely localized when $V_2/t>2$ (with the other chain still completely extended since $V_1=0$). 
As a result, in the special limit of $\tp=0$ and $V_2/t>2$, the system will be in an apparent SPIP, and both $\avgIPR$ and $\avgNPR$ are finite. 
This SPIP phase in the limit of $\tp =0$ is a pure mixed state and does not reflect an SPME.  
Thus, SPIP does not necessarily imply SPME although the existence of SPME is sufficient to produce SPIP. 

We now introduce a nonzero $\tp$ and study how the localization property of the system changes with $V_2$. 
In Fig.~\ref{Fig:CoupledChain1c} we plot $\avgIPR$ and $\avgNPR$ as a function of $V_2$ for two different $\tp$ values. 
When $\tp$ is small [Fig.~\ref{Fig:CoupledChain1c}(a)], the first localized states still appear around $V_2 = 2t$. 
In addition, when $V_2 > 2t$ both $\avgIPR$ and $\avgNPR$ are finite, indicating that the system still remains in the single-particle intermediate phase. 
However, it is interesting to note that even for very large $V_2$ values ($V_2\sim 10t$) the coupled system still cannot enter a regime where all eigenstates are localized. 
Therefore, the system hosts a rather wide single-particle intermediate phase. 
In contrast, when $\tp$ takes a larger value [Fig.~\ref{Fig:CoupledChain1c}(b)], there are two important changes. 
First, the first localized states in the energy spectrum now appear at a much larger $V_2$ value. 
Second, a completely localized phase is now able to emerge at large $V_2$ values, indicating that when the two chains are strongly hybridized, a large quasiperiodic potential on one chain (while the other is completely free of disorder) is sufficient to localize the coupled system. 
Here is a simple noninteracting example of a `bath' being localized by a system due to bath-system hybridization! 

\subsection{Phase diagram as a function of \texorpdfstring{$V_2$}{V} and \texorpdfstring{$\tp$}{tp}}

\begin{figure}[!]
\includegraphics[width = 3.5in]{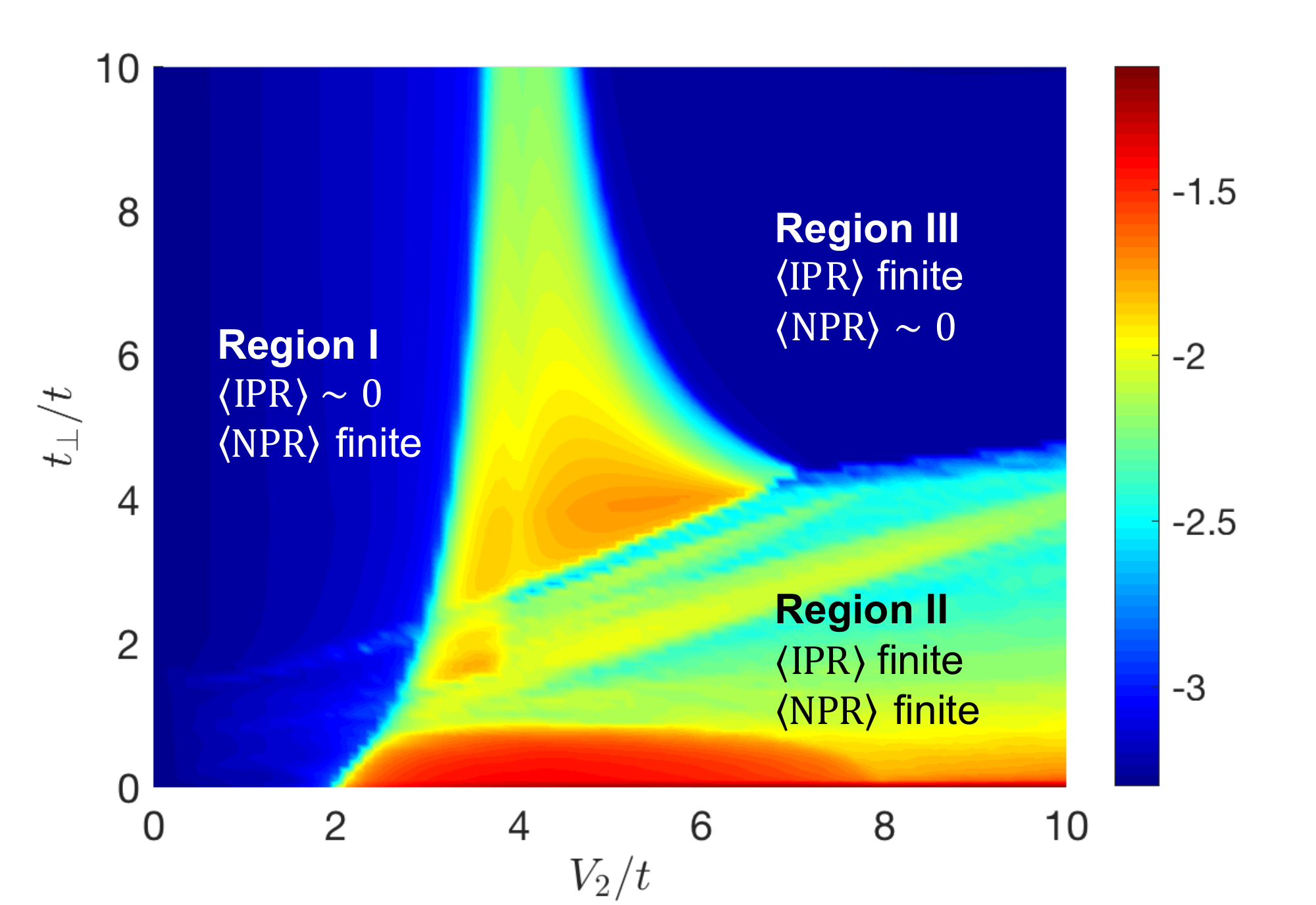}
\caption{\label{Fig:CoupledChainPhaseDiagram} Localization phase diagram in the coupled chain model. The color scale shows the quantity $\eta$ defined in Eq.~\eqref{Eq:eta}. 
 The size of each chain is $L=1000$, and $\alpha = 2/(\sqrt{5}-1)$.}
\end{figure}

Finally, we present the complete phase diagram of this coupled {\AAM} model in the $V_2$-$\tp$ plane. 
Our goal is to clearly identify the three distinct regions in the localization phase diagram of this model, as summarized in Table~\ref{Table:SPIP}. 
For this purpose we now introduce a new quantity $\eta$, defined as
\begin{align}
	\eta = \log_{10}\left[\avgIPR \times \avgNPR \right]. \label{Eq:eta}
\end{align}	
When both $\avgIPR$ and $\avgNPR$ are finite [$\sim \mathcal{O}(1)$] we have $-2 \lesssim \eta \lesssim -1$. 
In contrast, when either of them is $\sim L^{-1}$ (where $L$ is the size of the system) we have $\eta < -\log_{10}L$. 
For example, when $L\sim 10^3$, we have $\eta < -3$, which is distinct from the previous scenario. 
As a result, the quantity $\eta$ can help us clearly identify the single-particle intermediate phase in the phase diagram.

Figure~\ref{Fig:CoupledChainPhaseDiagram} presents the phase diagram in this coupled chain model by plotting the value of $\eta$. 
We can identify the three regions discussed above. In particular, we find that there is a rather wide intermediate phase (region II) in the phase diagram, especially for a small $\tp$. 
As $\tp$ becomes large, however, the size of the intermediate phase quickly shrinks. 
Another interesting feature in the phase diagram is that as $\tp$ increases the boundary between regions I and II shifts to larger values of $V_2$, indicating that it is more difficult for the first localized eigenstates to emerge in the energy spectrum as $\tp$ increases. 
The two-chain system with only two tunable system parameters ($\tp$ and $V_2$) is thus a rather rich noninteracting system to study the interplay between localized and extended states leading to an effective intermediate phase, ranging from a simple mixture phase  for $\tp=0$ and $V_2/t>2$ to complicated SPIP with nontrivial SPME for nonzero $\tp$ values.
Some additional details of our numerical results are given in Appendix~\ref{Appendix:A}. 

\begin{table}[!]
\begin{tabular}{c|c}
\hline\hline
(I) Extended phase & $\avgIPR\sim L^{-1}$ and $\avgNPR\sim\mathcal{O}(1)$\\\hline
(II) Intermediate phase (SPIP) & $\avgIPR\sim \mathcal{O}(1)$ and $\avgNPR\sim\mathcal{O}(1)$\\\hline
(III) Localized phase & $\avgIPR\sim \mathcal{O}(1)$ and $\avgNPR\sim L^{-1}$\\
\hline\hline
\end{tabular}
\caption{A convenient operational definition for the three different localization phases in a 1D single-particle Hamiltonian. Here $L$ is the size of the system. \label{Table:SPIP}}
\end{table}

\section{Discussions \label{Section:Discussion}}
\begin{figure}[!]
\includegraphics[width = 3.0in]{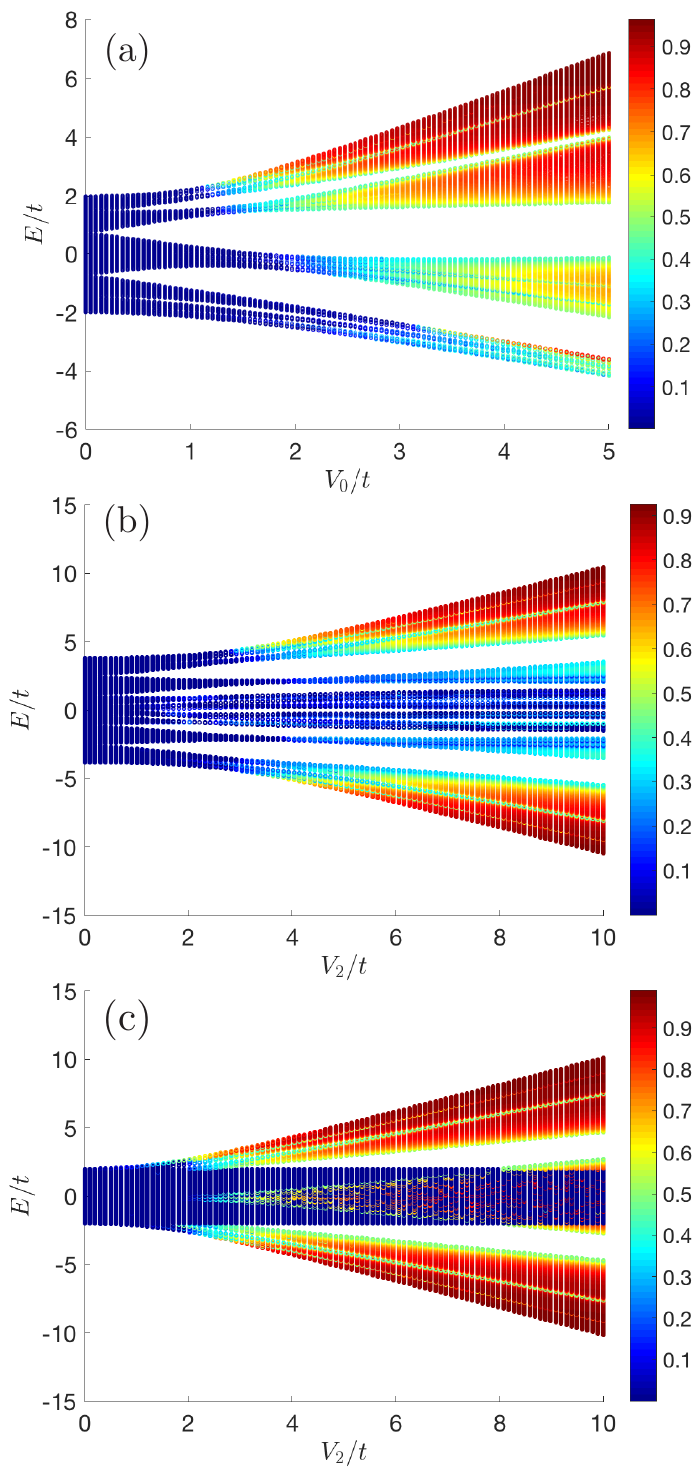}
\caption{\label{Fig:SPME} Illustration of the single-particle intermediate phase in (a) the {\SE} model and (b)-(c) the coupled {\AAM} model. The color bar in each figure shows the IPR value of the eigenstates. The system size is $L=10^3$ in all three figures. In addition, the parameters for the {\SE} model in (a) are $V_1=1/3$ and $Q=2/(\sqrt{5}-1)$, while the parameters for the coupled {\AAM} model are $\alpha = 2/(\sqrt{5}-1)$, $V_1=0$, with $\tp=1.80t$ in (b) and $\tp=0$ in (c).}
\end{figure}

Before we conclude, we make a few remarks on the relation between the single-particle mobility edge (SPME) and the single-particle intermediate phase (SPIP). 
We define the SPME as a critical energy $E_c$ in the energy spectrum which clearly separates localized and extended single-particle eigenstates. 
Furthermore, when approaching $E_c$ from the localized side of the energy spectrum, the Lyapunov exponent $\gamma(E)$ of the eigenstates will vanish as a power-law function of $E-E_c$, and a critical exponent for $\gamma(E)$ can be extracted according to Eq.~\eqref{Eq:ScalingExponent}. 
Meanwhile, we define the SPIP as the situation where the energy spectrum of a single-particle Hamiltonian contains both extended and localized states. 
A convenient operational definition of SPIP is the regime when both $\avgIPR$ and $\avgNPR$ are finite, as shown in Table~\ref{Table:SPIP}. 
SPME and SPIP are often used interchangeably in the literature, but we point out that they are not equivalent. 
Specifically, the existence of an SPME will always lead to the existence of an SPIP, but the reverse is not necessarily true. 
There can be SPIP without any SPME, thus making SPIP (as defined in Table~\ref{Table:SPIP} by both $\avgIPR$ and $\avgNPR$ being nonzero) a more general concept than SPME (as defined by a critical transition at a sharp mobility edge energy $E_c$ separating extended and localized states). 

To illustrate this point, we show in Fig.~\ref{Fig:SPME} three representative eigenstate IPR plots for the two models we have studied in this paper. 
In Fig.~\ref{Fig:SPME}(a) we can clearly see the existence of an SPME in the {\SE} model for $1.3 \lesssim V_0/t \lesssim 3.5$. In this regime, the localized states appear at high energies while the extended states reside in the low-energy part of the spectrum. 
In Fig.~\ref{Fig:SPME}(b) we find that when $\tp>0$ the energy spectrum of the coupled {\AAM} model also contains mobility edges. 
Actually we now find two different SPMEs, which divide the energy spectrum into a structure where extended eigenstates appear in the middle while the localized ones appear near the two edges. 
Incidentally, such a structure is exactly the same as that found in the well-known 3D Anderson model in the presence of random disorder.

By contrast, the $\tp=0$ limit of the coupled {\AAM} model shown in Fig.~\ref{Fig:SPME}(c)  is a clear departure from the above two examples. 
Although the system still consists of both extended and localized eigenstates (and thus the system is again in the SPIP), there is clearly no SPME in the energy spectrum. 
In fact, in the middle of the energy spectrum these two types of eigenstates are interwoven with each other, and they often reside at the same energy density. 
The physics behind this structure is rather simple: we are literally combining the localized states on one chain with the extended states on the other, without introducing any couplings between them. 
As a result, there is no level repulsion between these two types of eigenstates, giving rise to a completely mixed energy spectrum. 
The $\tp=0$ limit of the coupled {\AAM} model thus serves as a clear example of an SPIP without an SPME.
Another example of such an SPIP without any SPME is the $s$-$p$ model of Ref.~\cite{Xiao17}.  
Recently, a specific interacting  model  with an intermediate phase, but  without any mobility edge, between ETH and MBL has been proposed in Ref.~[\onlinecite{Schecter2018}]. 

Note that the distinction between an SPIP and an SPME can be made very sharp in solid-state systems, because there it is possible to detect the existence (or not) of an SPME by carrying out transport measurements. 
Specifically, due to the very high Fermi temperature in generic solid state materials, all transport measurements (even room-temperature ones) are effectively carried out in the zero-temperature limit. 
As a result, only states near the Fermi energy contribute to the transport. 
Consequently, when we measure the transport coefficients (such as conductance) as a function of the Fermi energy, we are able to map out an energy-resolved localization phase diagram for all the eigenstates. 
The familiar conductance plateau structure in the integer quantum Hall effects~\cite{Sarma1996} is a clear manifestation of the localization properties of eigenstates in each Landau level while the transitions between the plateaus occur through extended states at the centers of the Landau levels. 

{
In fact, we can make the following analogy between the three phases (extended, localized, and intermediate) in quasiperiodic models and our traditional notion of metals and insulators. 
For simplicity, we assume that there is only one SPME at $E_c$ in the energy spectrum, which separates localized states below $E_c$ from extended states above it. 
In this context, the purely extended phase ($\avgIPR\sim1$, $\avgNPR\sim0$) will be a pure metal at zero and finite temperatures $T>0$ with ballistic or diffusive transport. 
In contrast, the localized phase ($\avgIPR\sim0$, $\avgNPR\sim1$) is a pure insulator with zero conductivity even at finite 
temperatures. 
Finally, the intermediate phase ($\avgIPR$ and $\avgNPR$ both finite) is a system with insulating transport at finite $T$ but pure insulator at $T=0$ when the Fermi energy is at or below the SPME $E_c$. 
Specifically, the system will display no conductivity at $T=0$ but a finite yet exponentially small conductivity at finite $T$ and also reflecting strong system-to-system fluctuations in
conductivity. 
However, when the Fermi energy is above the SPME, it is difficult to distinguish the intermediate phase from the extended phase, as both systems will display ballistic or diffusive transport even at $T=0$. 
Thus, one can loosely think of the three phases as a pure metal (finite conductivity), pure insulator (zero conductivity), and exponentially small conductivity, respectively. 
}

By contrast, the distinction between an SPIP and an SPME is all but lost in the context of current ultracold atomic experiments, because an energy-resolved measurement is no longer feasible, for two reasons. 
On the one hand, a reliable zero-temperature transport-type measurement is currently lacking in ultracold atomic systems. 
All common experimental diagnostics such as time-of-flight measurements do not have an energy resolution. 
On the other hand, for MBL related experiments there is an additional complication. 
Because MBL is a dynamical phase of matter, a quench experiment is necessary to ascertain the long-time behavior of a given system. 
However, it is practically impossible to prepare a single eigenstate as the initial state, and thus all measurements in a quench experiment always involve a majority of the eigenstates. 
Therefore, strictly speaking, current quench experiments are only able to detect the existence of an SPIP but not an SPME. Only when we combine the experimental results with the theoretical knowledge of a model can we verify the existence of an SPME itself~\cite{LuschenSPME17}. 
We note that the mobility edge separates individual energy eigenstates whereas an intermediate phase involves all energy eigenstates, and thus any distinction between them must involve the ability to discern individual energy eigenstates. 

At this point, the distinction between an SPIP and an SPME seems abstract. 
After all, for a practical purpose, all we need is a system with an SPIP, and the two models in Fig.~\ref{Fig:SPME}(b)-(c) may be equally good candidates to study the interplay between localized and extended orbitals. 
We hereby draw the community's attention to such a distinction because it has profound implications when we generalize these two concepts to a many-body system. 
Specifically, the lesson we learn in this section is that the existence of an SPIP does not necessitate the existence of an SPME as a prerequisite. Therefore, these two concepts are not equivalent. 
When we generalize these arguments to a many-body system, it is thus plausible that the existence of a many-body intermediate phase (i.e., the coexistence of MBL and ETH states in the same system) is not contingent on the existence of a many-body mobility edge (a critical energy density that separates MBL and ETH states). 
Consequently, even if a many-body mobility edge does not exist in the thermodynamic limit in systems with random disorder~\cite{Roeck16,Roeck2017}, the existence of a many-body intermediate phase, in general, is not automatically ruled out. 
Instead, additional theoretical and experimental efforts are needed to answer this important question. 
In particular, energy resolved individual eigenstates must be studied in order to see the mobility edge whereas quench experiments starting from generic (i.e., non-energy-eigenstate) states can only probe the presence or absence of an intermediate phase without any information on whether an underlying mobility edge exists or not.

\section{Summary and Conclusion \label{Section:Summary}}

To summarize, in this work we have studied two one-dimensional models with a single-particle mobility edge as well as the single-particle intermediate phase that comes with it. 
We started with the Soukoulis-Economou model. 
We first showed that by choosing an appropriate value for the second incommensurate potential $V'$, this model can have a much wider single-particle intermediate phase than the one realized in recent experiments~\cite{LuschenSPME17}, thus providing a promising platform to study the coupling between localized and extended degrees of freedom in the same system. 
In addition, we studied the scaling behavior of Lyapunov exponent of the eigenstates near the single-particle mobility edge $E_c$, and evaluated the corresponding critical exponents. 
We further checked the same critical exponents in a family of incommensurate lattice models with a single-particle mobility edge, and found that this critical exponent depends sensitively on the details of the incommensurate potential, a behavior very different from that found in 3D Anderson models. 

The second model we have discussed is a quasi-1D model consisting of two 1D {\AAM} chains of the same size. 
In particular, we kept one chain completely extended (i.e., without any incommensurate potentials) while the other one subject to a finite incommensurate lattice potential with strength $V_2$. 
We described in detail how the localization properties of the coupled system depend on the inter-chain hopping $\tp$. 
There are two main results in this study. 
First, we mapped out the localization phase diagram as a function of $V_2$ and $\tp$, and found that there exists a very wide single-particle intermediate phase, especially for a small $\tp$. 
Second, we carefully analyzed how the inter-chain hopping $\tp$ affects the localization properties of the coupled system. 
Specifically, we find that at a small $\tp$ the coupling between the two chains delocalizes the localized orbitals on the disordered chain. 
In contrast, for a large $\tp$ the coupling between the two chains forces the extended orbitals on the clean chain to become localized. 

In addition, we also provide a detailed discussion on the role of the mobility edge in producing an intermediate phase with nonvanishing $\avgIPR$ and $\avgNPR$. 
We emphasize that while the existence of a mobility edge is sufficient to produce an intermediate phase, the reverse is not true, i.e., a mobility edge is not necessary in order to have an intermediate phase.

Given that the coupling between many-body localized states and ergodic states is a focus of current MBL research, we hope that the two models we discussed in this work can foster future experimental studies in this direction. 
In particular, with the existence of a wide single-particle intermediate phase and a relatively easy experimental implementation, these two models provide a promising platform to explore many-body effects on the single-particle mobility edge in one-dimensional systems.

\section*{Acknowledgment}
This work is supported by Microsoft and Laboratory for Physical Sciences.

\appendix

\section{Additional properties of the single-particle states in the two models \label{Appendix:A}}

In this Appendix we provide a few concrete properties of the two models we discuss in the main text, i.e., the {\SE} model and the coupled AA chain model. 
In particular, we will compare wave functions of eigenstates and the density of states in all three phases, i.e., localized, intermediate, and extended. 
Such a comparison will help us better understand these different regimes of single-particle states. 

\begin{figure*}[!]
\includegraphics[width = 6in]{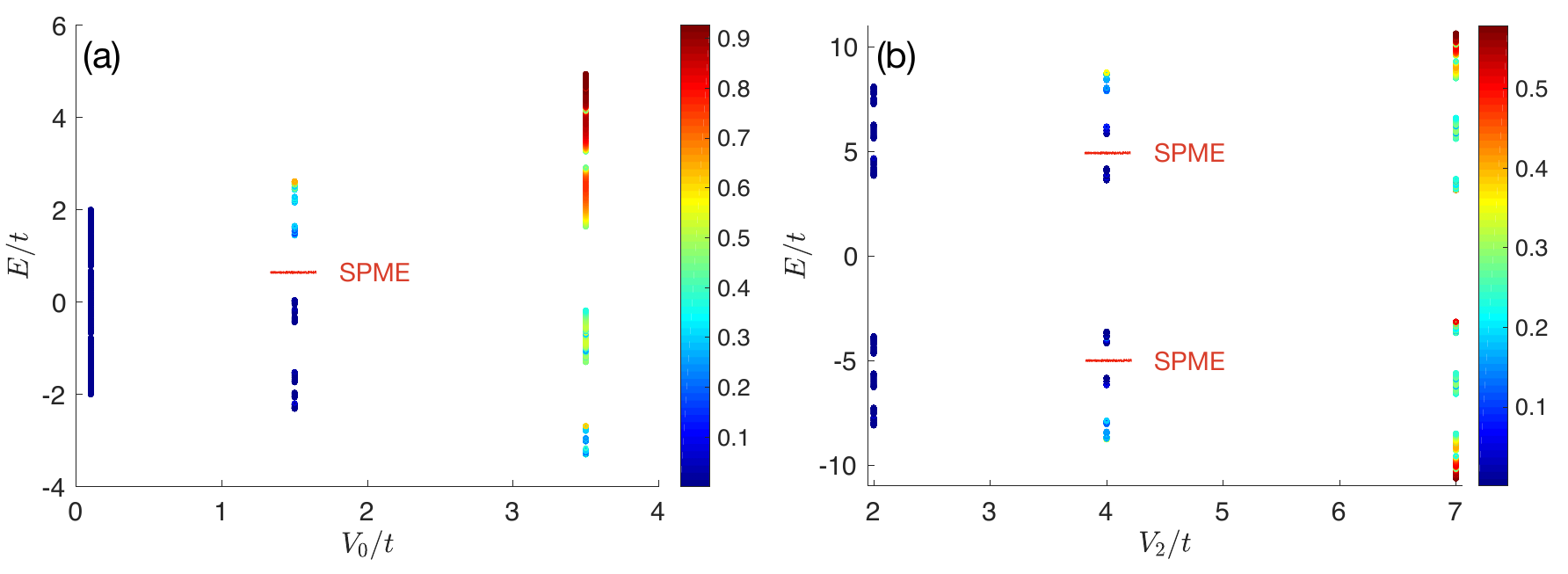}
\caption{\label{Fig:SPMEZOOM} 
 Illustration of the SPME in the (a) {\SE} model and (b) the coupled AA chain model. The color bar shows the IPR value for each eigenstate. 
 The parameters in (a) are: $L=10^3$, $V_1/V_0=1/3$, $\alpha=2/(\sqrt{5}-1)$, $\phi=0$, and the three $V_0$ values are $V_0/t = 0.1$, $1.5$, and $3.5$, respectively. 
 The parameters in (b) are: $L=10^3$, $\tp = 5.92$ (the same as Fig.~\ref{Fig:CoupledChain1c}), and $\alpha = 2/(\sqrt{5}-1)$. The three $V_2$ values are $V_2/t=2.0$, $4.0$, and $7.0$, respectively. 
}
\end{figure*}

\begin{figure*}[!]
\includegraphics[width = 6in]{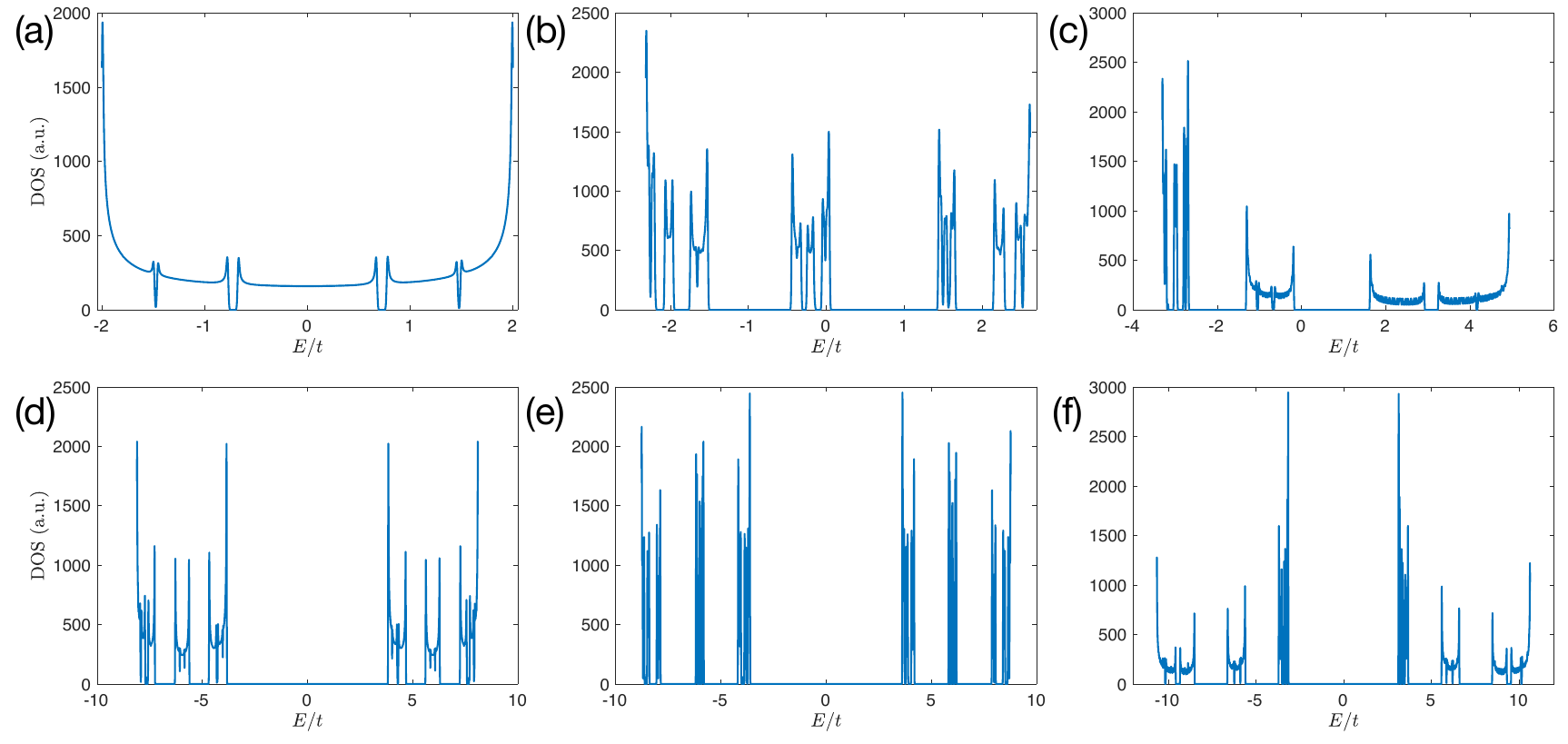}
\caption{\label{Fig:DOS} 
Density of states plot for (a)-(c) the {\SE} model and (d-f) the coupled AA chain model. The parameters for panels (a-c) are the same as panel (a) in Fig.~\ref{Fig:SPMEZOOM}, with $V_0/t=0.1, 1.5, 3.5$ in (a)-(c), respectively. 
Similarly the parameters for panels (d)-(f) are the same as panel (b) in Fig.~\ref{Fig:SPMEZOOM}, with $V_2/t = 2.0$, $4.0$, and $7.0$ in (d)-(f) respectively. 
}
\end{figure*}

In Fig.~\ref{Fig:SPMEZOOM} we show three typical energy spectra for each model, representing the localized, intermediate, and extended regimes, respectively. 
In particular, for the intermediate regime, we mark out the location of the SPME in the energy spectrum, which in this case lies in the energy gap. 
Moreover, note that there is only one SPME in the {\SE} model, while there are two SPMEs in the coupled AA chain model. 
Accompanying these energy spectra plots, we also present the corresponding density-of-states (DOS) plots in Fig.~\ref{Fig:DOS}. It can be seen clearly that as the strength of the quasiperiodic potential increases the spectrum splits into more and more subbands, which is a common feature for this type of incommensurate lattice models~\cite{Xiao17}. 

\begin{figure*}[!]
\includegraphics[width = 6in]{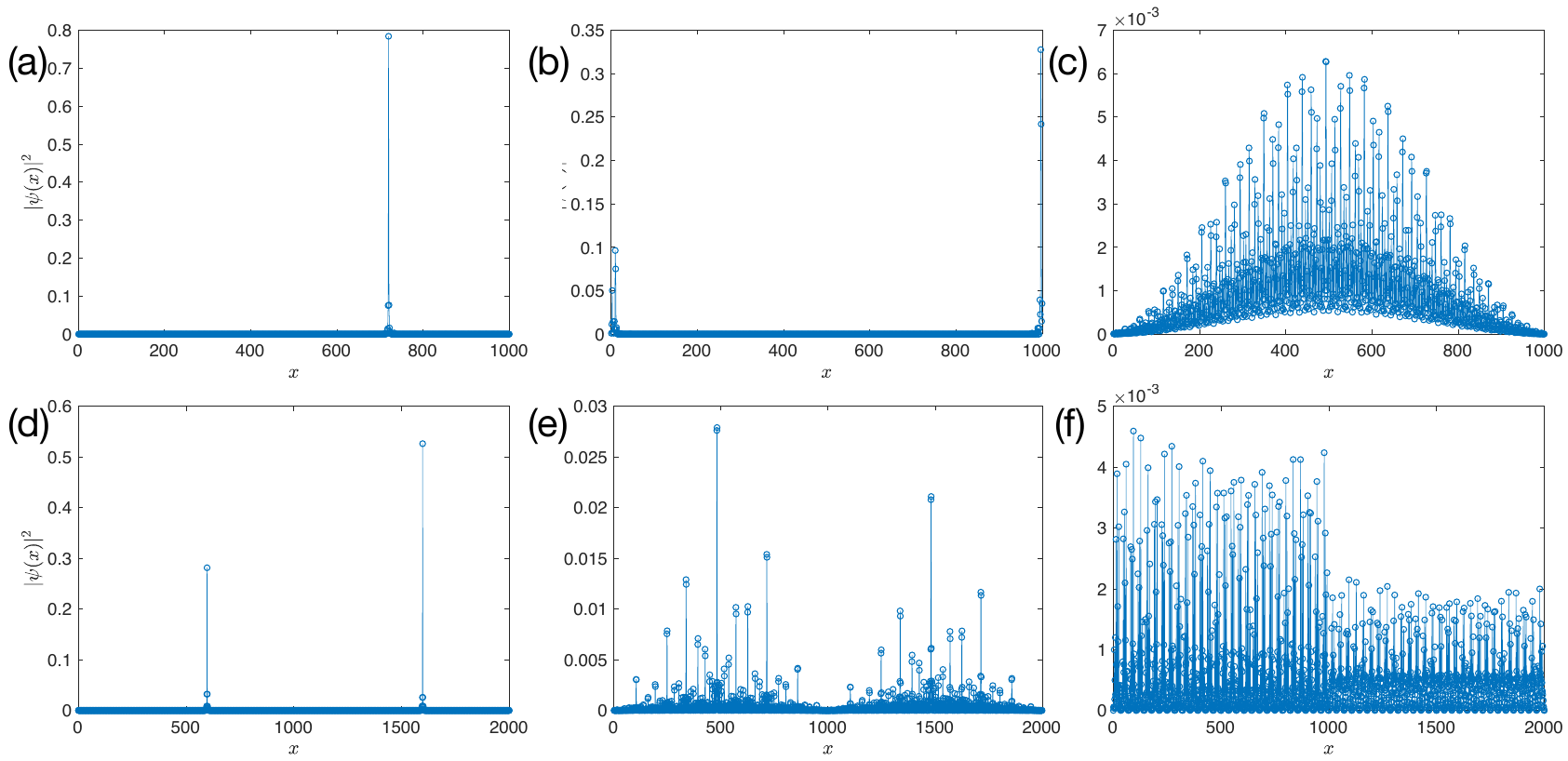}
\caption{\label{Fig:WF} 
(a)-(c) Wave function amplitudes for the {\SE} model. The parameters in this model are given by $L=10^3$, $V_1/V_0=1/3$, $\alpha = 2/(\sqrt{5}-1)$, $\phi=0$, and $V_0/t=1.5$. 
Moreover, (a)-(c) correspond to a localized, critical (the first state below the SPME), and extended eigenstate, respectively. 
(d)-(f) Wave function amplitudes for the coupled AA chain model. The parameters in this model are the following: $L=10^3$, $\tp=5.92$ (the same as Fig.~\ref{Fig:CoupledChain1c}), and $V_2/t=4.0$. 
The three panels (d)-(f) corresponds to a localized, critical (the first state above the SPME near $E=5t$ shown in Fig.~\ref{Fig:SPMEZOOM}), and extended eigenstate, respectively.  
}
\end{figure*}

In Fig.~\ref{Fig:WF} we show typical wave function amplitudes for states in the three regimes for these two models. 
It can be seen that localized states, extended states and states near the SPME (critical states) have rather different forms of wave functions. 
In particular, for eigenstates near the SPME, their wave function may look  localized in a local region, but there is a nonzero probability to find additional density peaks in other parts of the system. 
As a result, they are called critical states, and they can be shown to possess a fractal dimension~\cite{Xiao17}.

\section{Two duality relations \label{Appendix:B}}

In this Appendix we provide two results on the duality relations between certain incommensurate lattice models. 
The first one is to establish the duality between the {\SE} model~\cite{Soukoulis_Economou} and the $t_1$-$t_2$ model of Biddle et al.~\cite{SPME08}, which have been studied in this work. 
The second one is to derive the dual model to the long-ranged hopping model of Ref.~\cite{SPME07}, given by 
\begin{align}
	E u_n = t\sum_{m\neq n} e^{-a|n-m|} u_{m} + V\cos(2\pi\alpha n + \phi) u_n, \label{EqAppendix:LongRangeHopping}
\end{align}
where $a>0$ controls how fast the long-ranged hopping terms decay. 
In particular, in the $a\to +\infty$ limit the above model reduces to the familiar {\AAM} model. 

\subsection{Duality between the {\SE} model and the \texorpdfstring{$t_1$-$t_2$}{t1-t2} model}

The {\SE} model is given by 
\begin{align}
	E u_n = t(u_{n+1} + u_{n-1}) + \left[V_0\cos(Qn) + V_1\cos(2Qn)\right] u_n, \label{EqAppendix:SEModel}
\end{align}
where $n$ is the index for the lattice sites. 
We now introduce the Fourier transform of $u_n$, 
\begin{align}
	\tu_k = \dfrac{1}{\sqrt{L}} \sum_{n}e^{ikn}u_n, \; 
	u_n = \dfrac{1}{\sqrt{L}} \sum_{k} e^{-ikn}\tu_k, \label{EqAppendix:FourierTransform}
\end{align}
where $L$ is the size of the one-dimensional system. As a result, we can rewrite Eq.~\eqref{EqAppendix:SEModel} as 
\begin{align}
	E \sum_{k} e^{-ikn}\tu_k =& 2t \sum_{k}\tu_k e^{-ikn}\cos(k) \\ 
	&+ \sum_{k}\tu_k e^{-ikn}\left[V_0 \cos(Qn) + V_1\cos(2Qn)\right] \notag. 
\end{align}
We further multiply both sides by $e^{iqn}/L$ and sum over $n$, and obtain the following result, 
\begin{align}
	E\tu_q = 2t\cos(q) \tu_q + \dfrac{V_0}{2}(\tu_{q+Q} + \tu_{q-Q}) + \dfrac{V_1}{2}(\tu_{q+2Q} + \tu_{q-2Q}). \label{EqAppendix:FourierOfSE}
\end{align}

In order to establish the duality relation in a finite-size system, we need to approximate the irrational $Q$ by $Q = 2\pi M/L$, with $M$ and $L$ being coprime integers. 
Such an approximation allows us to further rescale $q$ to $m$ via the following relation: 
\begin{align}
	q = mQ~\mod~2\pi, m = 0, 1, 2, \ldots, L-1. \label{EqAppendix:Trick}
\end{align}
If we then define $\tu_q \equiv u_m$, we can finally rewrite Eq.~\eqref{EqAppendix:FourierOfSE} as 
\begin{align}
	Eu_m = 2t\cos(mQ)u_m + \dfrac{V_0}{2} (u_{m+1} + u_{m-1}) + \dfrac{V_1}{2}(u_{m+2}+u_{m-2}), 
\end{align}
which is exactly the $t_1$-$t_2$ model of Biddle et al.~\cite{SPME08}. 

\begin{widetext}
\subsection{The dual model of \texorpdfstring{Eq.~\eqref{EqAppendix:LongRangeHopping}}{Eq.(B1)}} 
We now derive the dual model of Eq.~\eqref{EqAppendix:LongRangeHopping}. 
We again use the Fourier transform in Eq.~\eqref{EqAppendix:FourierTransform} to rewrite Eq.~\eqref{EqAppendix:LongRangeHopping} as follows, 
\begin{align}
	E\sum_{k}\tu_k e^{-ikn} = V\cos(2\pi\alpha n+\phi) \sum_{k}e^{-ikn}\tu_k + t\sum_{k,m\neq n} e^{-ikm} \tu_k e^{-a|n-m|}. 
\end{align}
We further multiply both sides by $e^{iqn}/L$ and sum over $n$, and find that 
\begin{align}
	E\tu_{q} = \dfrac{V}{2}\biggl[e^{i\phi}\tu_{q+2\pi\alpha} + e^{-i\phi}\tu_{q-2\pi\alpha}\biggr] + \dfrac{1}{L}\sum_{n}e^{iqn} \sum_{k}\tu_k\left(t\sum_{m\neq n}e^{-ikm} e^{-a|n-m|}\right). \label{EqAppendix:FourierOfLongRangeHopping}
\end{align}
We now evaluate the second term in the above equation. 
Note that the terms in the round bracket can be written explicitly as 
\begin{align}
	t \sum_{m\neq n}e^{-ikm}e^{-a|n-m|} 
	=&t \biggl[ e^{-a} \left(e^{-ik(n+1)} + e^{-ik(n-1)}\right) 
	+ e^{-2a}\left( e^{-ik(n+2)} + e^{-ik(n-2)}\right)
	+ \cdots + e^{-sa}\left( e^{-ik(n+s)} + e^{-ik(n-s)}\right) + \cdots\biggr] \notag\\
	=&t_1 \biggl[ \left(e^{-ik(n+1)} + e^{-ik(n-1)}\right) 
	+ e^{-a}\left( e^{-ik(n+2)} + e^{-ik(n-2)}\right)
	+ \cdots + e^{-(s-1)a}\left( e^{-ik(n+s)} + e^{-ik(n-s)}\right) + \cdots\biggr], 
\end{align}
where we have identified $t_1\equiv te^{-a}$ as the nearest-neighbor hopping energy. 
Now insert the above result back to Eq.~\eqref{EqAppendix:FourierOfLongRangeHopping} and the entire second term there can be written explicitly as 
\begin{align}
	\dfrac{1}{L}\sum_{n}e^{iqn} \sum_{k}\tu_k\left(t\sum_{m\neq n}e^{-ikm} e^{-a|n-m|}\right) 
	=& \dfrac{2t_1}{L} \sum_{n,k} e^{i(q-k)n}\tu_k \biggl[ \cos(k) + e^{-a}\cos(2k) + \cdots + e^{-(s-1)a} \cos(sk) + \cdots\biggr] \notag\\
	=& 2t_1 \tu_q \biggl[ \cos(q) + e^{-a}\cos(2q) + \cdots + e^{-(s-1)a} \cos(sq) + \cdots\biggr] \notag. 
\end{align}
Putting everything together, we find that in Fourier space the long-ranged hopping model has the following form, 
\begin{align}
	E\tu_q = \dfrac{V}{2}\left[e^{i\phi}\tu_{q+2\pi\alpha} + e^{-i\phi}\tu_{q-2\pi\alpha}\right] + 2t_1 \tu_q \biggl[ \cos(q) + e^{-a}\cos(2q) + \cdots + e^{-(s-1)a} \cos(sq) + \cdots\biggr]. 
\end{align}
Given that this calculation is carried out in a finite-size system, we invoke the same trick as the one in Eq.~\eqref{EqAppendix:Trick}, and obtain 
\begin{align}
	Eu_m = \dfrac{V}{2}\left[e^{i\phi} u_{m+1} + e^{-i\phi}u_{m-1}\right] 
	+ 2t u_m \sum_{s} e^{-sa} \cos(2\pi sm\alpha), 
\end{align}
where we have restored the parameter $t$ in place of $t_1$. 
This is the dual model to the long-ranged hopping model given in Eq.~\eqref{EqAppendix:LongRangeHopping}. 
In particular, in the $a\to +\infty$ limit the above result and the long-ranged hopping model in Eq.~\eqref{EqAppendix:LongRangeHopping} are identical and equal to the {\AAM} model, which is a manifestation of the self-duality property of the {\AAM} model. 

\end{widetext}

\bibliography{Bib-SPME}
\end{document}